\documentclass[10pt,conference]{IEEEtran}
\IEEEoverridecommandlockouts
\usepackage{cite}
\usepackage{amsmath,amssymb,amsfonts}
\usepackage{algorithmic}
\usepackage{graphicx}
\usepackage{subfig}
\usepackage{standalone}
\usepackage{textcomp}
\usepackage{xcolor}
\usepackage{enumitem}
\usepackage{booktabs}
\usepackage{array}
\usepackage{tcolorbox}
\usepackage{multicol}
\usepackage{multirow}
\usepackage{makecell}
\usepackage{pgfplots}
\pgfplotsset{compat=newest}
\usepackage{adjustbox}
\usepackage{caption}
\usepackage{listings}

\usepackage[hyphens]{url}
\usepackage{breakurl}
\usepackage[hidelinks]{hyperref}

\usepackage[]{mdframed}
\usepackage{balance}
\usepackage[section]{placeins}
\usepackage{tabularx}
\usepackage{tablefootnote}
\def\BibTeX{{\rm B\kern-.05em{\sc i\kern-.025em b}\kern-.08em
    T\kern-.1667em\lower.7ex\hbox{E}\kern-.125emX}}
\begin{document}

\newcounter{findingcount}
\setcounter{findingcount}{1} 

\newtcolorbox{blockquote}{colback=white,colframe=black,fonttitle=\bfseries}

\newcommand{\revisioncolor}{black} 
\newcommand{\revision}[1]{{\color{\revisioncolor}#1}}

\lstdefinestyle{myStyle}{
    belowcaptionskip=1\baselineskip,
    breaklines=true,
    frame=none,
    numbers=none, 
    basicstyle=\footnotesize\ttfamily,
    keywordstyle=\bfseries\color{green!40!black},
    commentstyle=\itshape\color{purple!40!black},
    identifierstyle=\color{blue},
    backgroundcolor=\color{gray!10!white},
}

\newtcolorbox{quotebox}{colback=blue!10,boxrule=0.4pt,colframe=black,fonttitle=\bfseries,top=2pt,bottom=2pt}

\pgfplotsset{compat=1.11,
    /pgfplots/xbar legend/.style={
    /pgfplots/legend image code/.code={%
       \draw[##1,/tikz/.cd,yshift=-0.25em]
        (0cm,0cm) rectangle (3pt,0.8em);},
   },
}

\title{The Same Only Different: On Information Modality for Configuration Performance Analysis}

\author{
\IEEEauthorblockN{Hongyuan Liang$^{1}$, Yue Huang$^{1}$, Tao Chen$^{2 \ast}$}

\IEEEauthorblockA{$^1$ School of Computer Science and Engineering, University of Electronic Science and Technology of China, China}
\IEEEauthorblockA{$^2$ IDEAS Lab, School of Computer Science, University of Birmingham, United Kingdom}

\IEEEauthorblockA{lianghy16@outlook.com, huangyue16@outlook.com, t.chen@bham.ac.uk}

\thanks{$^{\ast}$Tao Chen is the corresponding author. Hongyuan Liang and Yue Huang are also supervised in the IDEAS Lab.}
}

\maketitle

\begin{abstract}
Configuration in software systems helps to ensure efficient operation and meet diverse user needs. Yet, some, if not all, configuration options have profound implications for the system's performance. Configuration performance analysis, wherein the key is to understand (or infer) the configuration options' relations and their impacts on performance, is crucial. Two major modalities exist that serve as the source information in the analysis: either the manual or source code. However, it remains unclear what roles they play in configuration performance analysis. Much work that relies on manuals claims their benefits of information richness and naturalness; while work that trusts the source code more prefers the structural information provided therein and criticizes the timeliness of manuals.

To fill such a gap, in this paper, we conduct an extensive empirical study over 10 systems, covering 1,694 options, 106,798 words in the manual, and 22,859,552 lines-of-code for investigating the usefulness of manual and code in two important tasks of configuration performance analysis, namely performance-sensitive options identification and the associated dependencies extraction. We reveal several new findings and insights, such as it is beneficial to fuse the manual and code modalities for both tasks; the current automated tools that rely on a single modality are far from being practically useful and generally remain incomparable to human analysis. All those pave the way for further advancing configuration performance analysis.


\end{abstract}


\section{Introduction}

Software systems are often designed with a daunting number of configuration options. For example, until now, \textsc{MySQL} has more than 600 configuration options that can be set by users~\cite{mysql}. Although this certainly indicates the great flexibility of the software, it also comes with a cost: it has been reported that 59\% of the performance issues worldwide are caused by configuration~\cite{DBLP:conf/esem/HanY16}. Therefore, much work has been done to mitigate these issues, mainly related to the modeling~\cite{10.1145/3702986,DBLP:conf/icse/HaZ19,DBLP:journals/pacmse/Gong024,DBLP:conf/sigsoft/Gong023,DBLP:journals/tse/GongChen25}, testing~\cite{DBLP:conf/kbse/HeJLXYYWL20,DBLP:conf/icse/MaChen25,DBLP:conf/issta/ChengZMX21}, and tuning~\cite{DBLP:conf/sigsoft/0001L21,DBLP:journals/tse/Nair0MSA20,DBLP:journals/pacmse/0001L24,DBLP:journals/tse/ChenCL24,DBLP:conf/icse/YeChen25,DBLP:journals/tosem/ChenL23a} of configuration.

However, all those research topics rely on or can benefit from two prerequisite tasks: (1) since not all the options are performance-sensitive, it is essential to preselect which are the ones that should be considered. For example, options that set a directory path or a port often do not impact performance. In contrast, options that control the code logic or some resource-heavy operations are performance sensitive, such as the \texttt{wait\_timeout} in \textsc{MySQL} and \texttt{maxThreads} in \textsc{Tomcat}; (2) there might be a dependency between two options, for example, in \textsc{Nginx}, the option \texttt{TCP\_NOPUSH} can be used only when option \texttt{sendfile} is enabled. As a result, it is fundamental to know these dependencies beforehand. Those prerequisite tasks are the key steps in configuration performance analysis~\cite{DBLP:journals/tse/SayaghKAP20,DBLP:conf/sigsoft/ChenCWYHM23,DBLP:conf/sigsoft/ChenWLLX20}. Indeed, much work has been done to investigate and propose tools for inferring performance-sensitive options~\cite{DBLP:conf/sigsoft/ChenCWYHM23,DBLP:conf/icse/HeJLYZ0WL22} and their dependencies~\cite{DBLP:conf/sosp/XuZHZSYZP13,DBLP:journals/pvldb/LaoWLWZCCTW24,DBLP:conf/sigsoft/ChenWLLX20}, but they leverage a single modality---either using manual (e.g., official guidelines or API docs) or the source code.


The motivation for using either modality in configuration performance analysis is arguable: on the one hand, researchers who are in favor of manuals state that these artifacts provide strong and intuitive domain knowledge with rich information to be extracted~\cite{DBLP:journals/pvldb/LaoWLWZCCTW24,DBLP:conf/icse/HeJLYZ0WL22}. On the other hand, work that relies on source code claims that it offers comprehensive structure and state of the systems~\cite{DBLP:conf/sigsoft/ChenCWYHM23,DBLP:conf/sigsoft/ChenWLLX20}, while being robust to the timeliness issue of the manual, i.e., what has been changed in the codebase might not have been timely reflected in the manual~\cite{DBLP:journals/csur/XuZ15,DBLP:journals/tse/SayaghKAP20}. Yet, there has been a lack of understanding on which modality should be preferred or tends to be more useful in configuration performance analysis and why.

In this paper, we seek to fulfill this gap. To that end, we conduct an extensive empirical study over 10 systems that come with a total of 1,694 configuration options, 106,798 words of manual, and 2,859,552 lines-of-code, aiming to understand the role of manual and code for identifying performance-sensitive options and extracting their dependencies in configuration performance analysis. We do that in two phases: in the first phase, we carry on a human analysis involving all authors that independently parse and analyze the manual and code, from each of which the results are compared. In this second phase, we examine the effectiveness of existing automated tools that leverage manual or code individually, while comparing their outcomes against those from the human analyzers. \revision{Through those, we aim to answer several important research questions (RQs) related to configuration performance analysis:}


\begin{itemize}
    \item \revision{\textbf{RQ1:} How can manual or code be relatively useful in identifying performance-sensitive options?}
    \item \revision{\textbf{RQ2:} Comparatively, to what extent does manual or code help to extract performance-sensitive dependencies?}
    \item \revision{\textbf{RQ3:} How do existing automated tools in configuration performance analysis perform compared with human analysis using either manual or code information?}
\end{itemize}





\revision{Among others, our key findings/contributions are that:}
\begin{itemize}
    \item To identify performance-sensitive options, the manual leads to fewer false negatives while the code is more false-positive resistant, but fusing both modalities can maximize the benefit. They both cause false positives due to misleading information but incur different causes for false negatives: the vague description in the manual and the non-standard patterns of executions in code.
    \item While code is often more helpful for extracting dependencies than manual, the manual can still be useful for some edge cases. We found no false positives for both modalities, and the false negatives of finding dependencies from manual and code are mainly due to lack of information and excessive recursions of function calls, respectively.
    \item Existing automated tools for both tasks are far from being practically useful, and they are not comparable to human analysis, except for identifying performance-sensitive options from the manual.
\end{itemize}

Deriving from these, we articulate several insights for this direction of research, such as the necessity of fusing both manual and code modality in identifying performance-sensitive options and the extraction of dependencies therein should also consider performance-insensitive options.

As part of our results, we have also made the dataset, code, and tools used available at: \textcolor{blue}{\url{https://github.com/ideas-labo/cp-mod}}. Compared with existing datasets~\cite{DBLP:journals/pvldb/LaoWLWZCCTW24,DBLP:conf/sigsoft/ChenCWYHM23,DBLP:conf/sigsoft/ChenWLLX20}, our labeled dataset is of a larger scale with more diverse systems, complementing the research for automated configuration performance analysis.

The reminder of this paper is as follows: Section~\ref{sec:bg} presents the background and motivation. Section~\ref{sec:methodology} delineates our empirical methodology. Section~\ref{sec:results} presents and analyzes the results followed by the insights learned in Section~\ref{sec:insights}. Sections~\ref{sec:threats} and~\ref{sec:related} present the threats to validity and related work, respectively. Section~\ref{sec:con} concludes the work.

\section{Background and Motivation}
\label{sec:bg}

\subsection{Analyzing Configurable Systems}

A configurable system can be configured by giving a vector of configuration values $\mathbf{c}=\{x_1,x_2,...,x_n\}$, where $x_n$ denotes the value of the $n$th configuration option, which can be binary, enumerate, or numeric~\cite{DBLP:conf/sigsoft/XuJFZPT15}. Configuration performance analysis studies the correlation between the configuration options and the system's performance, e.g., runtime or throughput, along with the interrelationships between the options. The results are fundamental for many configuration-related tasks, such as configuration performance modeling~\cite{DBLP:conf/icse/HaZ19,DBLP:conf/msr/GongC22}, configuration tuning/testing~\cite{DBLP:journals/tse/ChenCL24,DBLP:conf/kbse/HeJLXYYWL20}, and self-adaptation~\cite{DBLP:conf/wcre/Chen22,DBLP:conf/icse/Kumar0BB20,DBLP:conf/seams/Chen22,DBLP:journals/computer/ChenB15}.


\begin{figure}[t!]
\centering
\subfloat[Manual from \textsc{MySQL}]{ 


\adjustbox{max width=0.5\columnwidth}{
\begin{tabular}{|p{5.7cm}|}
 \toprule


\textbf{Option:} \verb|range_optimizer_max_mem_size| \\
\textbf{Partial Description:} 
The \verb|range_optimizer_max_mem_size| controls the limit on memory consumption for the range optimizer. A value of 0 means “no limit.” If an execution plan considered by the optimizer uses the range access method but the optimizer estimates that the amount of memory needed for this method would exceed the limit, it abandons the plan.\\

\bottomrule
\end{tabular}
}
}
~\hfill
\subfloat[Code from \textsc{MySQL}]{ 


\adjustbox{max width=0.5\columnwidth}{
\begin{tabular}{|p{5.85cm}|}
 \toprule

\textbf{Option:} \verb|select_into_buffer_size| \\
\textbf{Partial Code Snippet:} 
\begin{lstlisting}[language=C,style=myStyle,numbers=none]
if (init_io_cache(cache, file, thd->variables.select_into_buffer_size, WRITE_CACHE, 0L, true, MYF(MY_WME))) {
    mysql_file_close(file, MYF(0));
    mysql_file_delete(key_select_to_file, path, MYF(0));
    return -1;
  }
\end{lstlisting}\\

\bottomrule
\end{tabular}
}
}

\caption{Examples of manual texts and source code for revealing performance-sensitive options.}
\label{fig:exp1}
\vspace{-0.3cm}
\end{figure}

\subsection{Performance-sensitive Options}

Not all configuration options are performance-sensitive, therefore it is essential to identify which are performance-sensitive in configuration performance analysis. In essence, two common modalities exist for finding performance-sensitive options: the documentation manual and source code. The former often provides clear information about the performance sensitivity. As can be seen in \textcolor{black}{Figure~\ref{fig:exp1}a}, the description states that the option can control the allocated memory, which would clearly be important for the performance.

The latter, in contrast, contains performance-sensitive information with the code execution logic. For example, in \textcolor{black}{Figure~\ref{fig:exp1}b}, the call logic shows that \texttt{select
\_into\_buffer\_size} affects the buffer size for initializing the cache---a key performance indicator. 


\begin{figure}[t!]
\centering
\subfloat[Manual from \textsc{Nginx}]{ 


\adjustbox{max width=0.5\columnwidth}{
\begin{tabular}{|p{6.85cm}|}
\toprule
\textbf{Dependency:} \verb|sendfile| and \verb|TCP_NOPUSH| \\
\textbf{Partial Description:}\\
 In this configuration \texttt{sendfile} is called with the SF\_NODISKIO flag which causes it not to block on disk I/O, but, instead, report back that the data are not in memory. \textsc{Nginx} then initiates an asynchronous data load by reading one byte. On the first read, the FreeBSD kernel loads the first 128K bytes of a file into memory, although next reads will only load data in 16K chunks. This can be changed using the \texttt{read\_ahead}.\\
Enables or disables the use of the \texttt{TCP\_NOPUSH} socket option on FreeBSD or the \texttt{TCP\_CORK} socket option on Linux. The options are enabled only when \texttt{sendfile} is used.\\
\\

\bottomrule
\end{tabular}
}
}
~\hfill
\subfloat[Code from \textsc{HDFS}]{ 


\adjustbox{max width=0.5\columnwidth}{
\begin{tabular}{|p{7.2cm}|}
 \toprule
\textbf{Dependency:} \verb|dfs.replication| and \texttt{dfs.na}\\
\texttt{menode.maintenance.replication.min}\\
\textbf{Partial Code Snippet:} 
\begin{lstlisting}[language=C,style=myStyle,numbers=none]
final int minMaintenanceR = conf.getInt(
    DFSConfigKeys.DFS_NAMENODE_MAINTENANCE_ REPLICATION_MIN_KEY,DFSConfigKeys.DFS_NAMENODE_MAINTENANCE_
REPLICATION_MIN_DEFAULT);
......
    this.defaultReplication = conf.getInt(DFSConfigKeys.DFS_REPLICATION_KEY,DFSConfigKeys.DFS_REPLICATION_DEFAULT);
......
if (minMaintenanceR > defaultReplication) {
throw new IOException
......
\end{lstlisting}\\
\bottomrule
\end{tabular}
}
}

\caption{Examples of manual texts and source code for revealing performance-sensitive dependencies.}
\label{fig:exp2}
\vspace{-0.3cm}
\end{figure}

\subsection{Performance-sensitive Dependencies}


There can be a dependency between a pair of options. \revision{For example, in the \textsc{Hadoop} implementation of \textsc{MapReduce}, any value of the option \texttt{mapreduce.jobhistory.loadedjobs.cache.size} would be ignored if the option \texttt{mapreduce.jobhistory} \texttt{.loaded.tasks.cache.size} is set to a positive value.} Failure to comply with the dependency might result in, e.g., ineffective configuration or runtime exceptions.

Similar to the performance-sensitive options, extracting dependencies sensitive to the performance can also be achieved via analyzing the manual or code. Figure~\ref{fig:exp2}a shows an example of the manual, in which we see that the texts directly state that \texttt{TCP\_NOPUSH} depends on the value of \texttt{sendfile}. In contrast, the source code also provides such information through some joint implication to the control flow between the two options. For example, in Figure~\ref{fig:exp2}b, the two options control two variables independently, which would then be compared to determine if there is an exception, hence a dependency.

\revision{However, manually parsing/analyzing dependencies from the software manuals and code is a labor-intensive process (regardless whether they are performance-sensitive or not). This is because the description in the manual can be lengthy and hard to be interpreted; while the code of relevant options can involve complex call chains and code structure.}


\subsection{Motivation}

\subsubsection{Why Performance?}

\revision{It is worth noting that even valid options might not be performance-sensitive. Those non-performance-sensitive options overcomplicate tasks related to configuration performance, e.g., configuration tuning and debugging. Therefore, a common way in those works would be to select (either manually or automatically) only the performance-sensitive options to work on \cite{DBLP:conf/hotstorage/KanellisAV20,DBLP:conf/icse/HeJLYZ0WL22,DBLP:conf/kbse/ZhuH23,DBLP:journals/tse/ChenGongChen25}. As a result, it is crucial to specifically study performance-sensitive options and understand how their dependencies can impact the mutation of those options in the tuning/testing process, which is the key motivation of this work.}

\subsubsection{Why Manual and Code?}

Indeed, much work relies on either manual or code individually. Those who use manuals claim that the manual contains a wide range of natural descriptions of the options and, hence, is more information-rich~\cite{DBLP:journals/pvldb/LaoWLWZCCTW24,DBLP:conf/icse/HeJLYZ0WL22}. Conversely, work that prefers code analysis states that the code is more structural and is less prone to outdated issues caused by rapid version updating~\cite{DBLP:conf/sigsoft/ChenCWYHM23,DBLP:conf/sigsoft/ChenWLLX20,DBLP:journals/csur/XuZ15,DBLP:journals/tse/SayaghKAP20}. Despite the above, it remains unclear which modalities are more useful, particularly for identifying the performance-sensitive options and their dependencies. \revision{For example, \texttt{SafeTune}~\cite{DBLP:conf/icse/HeJLYZ0WL22} is an automated tool that can be used for configuration performance analysis, which only relies on manual texts while completely ignoring code information without giving clear justification}. This work seeks to bridge such a gap through an empirical study, providing insights into the roles of different modalities in configuration performance analysis.


\section{Methodology}
\label{sec:methodology}




\begin{table}[t!]
\centering

\setlength{\tabcolsep}{1mm}
\caption{Configurable software and the manuals studied.}

\label{tb:software}
\adjustbox{max width=\columnwidth}{
\begin{tabular}{lllrrrl}
 \toprule

\textbf{Software} & \textbf{Domain} & \textbf{Language} & \textbf{Words} & \textbf{LoC} & \textbf{Version} & \textbf{Used By} \\

\midrule
\textsc{Httpd} & Web Server & C & 6,161 & 634,144 & 2.4.57 &  \cite{DBLP:conf/kbse/WangHLLJJLL23}\cite{DBLP:conf/icse/HeJLYZ0WL22}\cite{DBLP:conf/qrs/ZhouLLDLX16}\\
\textsc{Lighttpd} & Web Server & C & 686 & 252,192 & 1.4 & \cite{DBLP:conf/qrs/ZhouLLDLX16} \\
\textsc{Nginx} & Web Server & C & 4,741 & 413,976 & 1.24.0 & \cite{DBLP:conf/kbse/WangHLLJJLL23}\cite{DBLP:conf/icse/HeJLYZ0WL22}\cite{DBLP:conf/qrs/ZhouLLDLX16}\\
\textsc{Redis} & In-memory Database & C & 14,614 & 437,076 & 6.2.12 & \cite{DBLP:conf/qrs/ZhouLLDLX16}\cite{DBLP:journals/jss/CaoBWZLZ23} \\
\textsc{MySQL} & SQL Database & C/C++ & 36,424 & 10,332,544 & 8.0.33 & \cite{DBLP:journals/pvldb/LaoWLWZCCTW24}\cite{DBLP:conf/sosp/XuZHZSYZP13} \\
\textsc{HDFS} & File System & Java & 7,974 & 2,344,750 & 3.3.5 & \cite{DBLP:conf/sigsoft/ChenWLLX20}
\cite{DBLP:journals/tse/ZhouHXJLWXWLB23}\\
\textsc{MapReduce}\tablefootnote{\revision{We use the \textsc{Hadoop} implementation of \textsc{MapReduce}.}} & Distributed Computing & Java & 7,677 & 708,280 & 3.3.5 & 
\cite{DBLP:conf/icse/HeJLYZ0WL22}\cite{DBLP:conf/sigsoft/ChenWLLX20}\cite{DBLP:conf/eurosys/LiWHL20}
\\
\textsc{HBase} & Distributed NoSQL Database & Java & 9,666 & 3,960,100 & 2.5.5 & \cite{DBLP:conf/sigsoft/ChenWLLX20}\cite{DBLP:journals/tse/ZhouHXJLWXWLB23}\cite{DBLP:conf/eurosys/LiWHL20}\\
\textsc{Yarn} & Resource Management & Java & 13,277 & 2,575,072 & 3.3.5 & \cite{DBLP:conf/sigsoft/ChenWLLX20}\cite{DBLP:journals/tse/ZhouHXJLWXWLB23}
\\
\textsc{Tomcat} & Application Server & Java & 9,498 & 1,201,418 & 10.1.16 & \cite{DBLP:conf/cloud/ZhuLGBMLSY17} \\
\midrule
\end{tabular}
}
\end{table}

\subsection{Software Systems}


To collect data from widely-used configurable systems, we study those open-sourced ones that also come with well-documented manuals. The list has been illustrated in Table~\ref{tb:software}. We can clearly see that the systems studied come from diverse domains, with different languages and scales, while being widely used by the community. To ensure the validity of each system, we take the latest stable version and analyze its code and the corresponding manual by the date of analysis, i.e., Sep 2023, representing the most stable state of a system since the newest releases are likely to be error-prone.

\subsection{Identifying Performance-sensitive Options}

\subsubsection{Manual Analysis}

Two authors who are experienced software engineers independently analyzed the manual to identify performance-sensitive options. This follows a mix of blacklist and whitelist approaches. The blacklist contains several keywords that, if detected in the manual's description, would cause the corresponding option to be eliminated immediately. In contrast, if a description contains any of the terms in the whitelist, then the corresponding option is considered performance-sensitive. The procedure is as follows:

\begin{enumerate}[label=(\alph*)]
    \item \textbf{Screening:} As from Table~\ref{tb:keyword}, we conduct an initial screening by identifying the keywords for blacklist (e.g., names and paths) and whitelist (e.g., resource and thread). 
    
    \item \textbf{Blacklist Exclusion:} We eliminate any options which have descriptions that contain the keywords in the blacklist. The two authors exchanged their filtered list and any disagreement would be discussed or consulting the other author or external expert until a consensus can be made.
    \item \textbf{Whitelist Inclusion:} In the remaining options, the description of which contains any of the terms in the whitelist is said as performance-sensitive. Extensive discussions were taken place to resolve disagreements. 
    \item The two authors independently read the options collected and picked the performance-sensitive ones as understood from the description. A final list is shared and discussed until agreements are reached. The same process is repeated for the manual of every system studied.
\end{enumerate}

\begin{table}[t!]
\centering

\caption{Keywords for extracting performance-sensitive options from the manual (all are case-insensitive).}

\label{tb:keyword}
\adjustbox{max width=\columnwidth}{
\begin{tabular}{lp{8cm}}
 \toprule

\textbf{} & \textbf{Terms}  \\

\midrule
Blacklist&
\textit{file, proxy, forward, path, port, address, location, updates, version}
\textit{compatibility, legacy, address, link, restore, plugin, dir, url, host, name, precision}
\textit{descriptor, principal, key, ui, profile, info, catalogue, password, pwd}\\

\hline

Whitelist&
\textit{debug, optimize, table, cpu, pct, interleave, block}
 \textit{thread, worker, executor, error, time, depth, max, logger, range}
  \textit{size, min, length, timeout, limit, cache, mode, log}\\
\bottomrule
\end{tabular}
}
\end{table}

For example, \texttt{mapreduce.cshuffle.port} is an option of \textsc{MapReduce} that was filtered out using the blacklist since it merely specifies the port for connecting the component. Conversely, for \textsc{MySQL}, the \texttt{wait\_timeout} is an option that is included under the whitelist since the waiting time would have a dramatic impact on the system's performance. 



\subsubsection{Code Analysis}
To understand the performance-sensitive options from code, we leverage a semi-automated approach that relies on taint analysis tools\footnote{\revision{Our taint analysis is technically similar to prior work~\cite{DBLP:conf/icse/VelezJSAK21,DBLP:journals/ase/VelezJSSAK20,DBLP:conf/ppopp/CopikCGW0H21}, but applied under different definitions of source and sink that fulfill our needs.}}. The same two leading authors took the lead, but all authors helped to resolve conflicts:

\begin{enumerate}[label=(\alph*)]
    \item \textbf{Operation Categorization:} Any operation/function that belongs to the categories below are performance sensitive:

    \begin{itemize}
        \item Memory operations, including array or stack allocations, e.g., the \texttt{init\_mem()} in \textsc{MySQL}. 
        \item Cache operations, updates or synchronization, e.g., the \texttt{ngx\_open\_cached\_file()} in \textsc{Nginx}.
        \item I/O operations, e.g., those accessing the persistent storage, such as the \texttt{BufferedReader} for JAVA.
        \item Multi-threading process, such as thread pooling, e.g., the \texttt{newCachedThreadPool} in \textsc{MySQL}.
        \item Loop operations, such as \texttt{for}, \texttt{while}, and \texttt{do-while} loops.
        \item Network communication: such as the \texttt{http\_range\_header\_filter()} in \textsc{Httpd}.
        \item Error control process: code such as \texttt{try-catch}, \texttt{assert}, \texttt{my\_error()}.
    \end{itemize}

    Those categories are discussed among all authors until an agreement has been reached.

   \item \textbf{Variable Identification:} It is easy to understand what the configuration options are by reading the configuration file, e.g., \texttt{yaml} files. Yet, since the configuration options in the code are codified in different ways depending on the systems, we use different approaches below to establish the mappings between configuration options from the manual and the corresponding variables in the code:
    \begin{itemize}
        \item \textbf{Unified Analysis}: For \textsc{MySQL}, \textsc{Lighttpd}, \textsc{Nginx}, and \textsc{Redis}, the variables for configuration options are centrally organized in a single file with identical names, and hence the mappings can be found directly.
        \item \textbf{Segmented Analysis}: For \textsc{Httpd} and \textsc{Tomcat}, the variables are written in separate source code files but they are still centrally involved in a \texttt{main} thread. For those, we analyze the control flow to find the relevant code files and the corresponding variables. 
        \item \textbf{Scattered Analysis}: For \textsc{HDFS}, \textsc{MapReduce}, \textsc{HBase}, and \textsc{Yarn}, the variables are distributed across the codebase and are only called on-demand. Yet, those variables are all accessible via the \texttt{setter()} and \texttt{getter()} function, which can be easily allocated. 
        
    \end{itemize}
   Again, the two lead authors conduct the analysis independently and exchange the results to reach a consensus.

    \item \textbf{Taint Analysis:} To extract those performance-sensitive options by analyzing the code, we leverage taint analysis, where we use variables identified serve as the sources and the performance-sensitive operations are the sink. For C/C$++$ systems, we use \texttt{Clang} \cite{llvm}, and for JAVA systems, we use \texttt{JavaParser} \cite{javaparser}, in two ways:

    \begin{itemize}
        \item\textcolor{black}{In the \textbf{top-down manner}, the taint analysis would procedure some taint flows that indicate which performance-sensitive operations would interact with the variables identified. However, it is unclear whether those variables can impact the behavior of the operation. To confirm that, we further extract the corresponding code fragments from the Abstract Syntax Tree to investigate whether the path between the variables identified and performance-sensitive operations involve logical operators, e.g., \texttt{If/Else} or \texttt{While}. If it does, we find a performance-sensitive option/variable.}

        \item \revision{We also follow a \textbf{bottom-up} approach starting from each of the performance-sensitive operations, based on which we identify the related parameters. We then manually traced back to the origin of those parameters to confirm their interactions with the variables. If an interaction exists, the corresponding variable is performance-sensitive. This helps to mitigate the false negative cases from the taint analysis.}
    \end{itemize}
\end{enumerate}

\subsubsection{System Measurement (Ground Truth)}

To identify the ground truth, we manually change each configuration option and examine whether such a change leads to significant performance deviation. \revision{This is achieved by running and profiling each system under 3-4 different benchmarks/workloads setting according to the commonly used ones from the literature~\cite{DBLP:journals/pvldb/ShiZLCLW14,DBLP:journals/jss/CaoBWZLZ23,DBLP:conf/icse/HeJLYZ0WL22,DBLP:conf/hotstorage/KanellisAV20} and practical applications \footnote{Details: \textcolor{blue}{\texttt{https://github.com/ideas-labo/cp-mod/blob\break /main/dataset/perfsensitive-groundtruth/workloads.md}}}. For example, we use \textsc{Sysbench} and \textsc{TPC} for \textsc{MySQL} with different concurrent users, test duration, and table counts, etc.}


For each option, we sample at least two possible values as the representatives: for the numeric options, we use the default, maximum, and minimum within the range documented. If no range is provided, we use the positive upper limit as the maximum value and 0 as the minimum value. For boolean options, we simply flip their \texttt{true} and \texttt{false}. For other enumerate options, we compare all pairs of the possible values in the eligible range. For all cases, we compare the performance achieved by setting the two values of an option, while keeping the other options with their default values.


We repeat the process five times to ensure the stability and reliability. A numeric or boolean option is said to be indeed performance-sensitive if, under any workload, the average performance of setting one value can have more than $5\%$ deviation compared with that of the setting the said option with the other value; an enumerate option is performance-sensitive if the average performance of any pair of its value has deviation greater than $5\%$ under any workload. The above is a typical setting being acknowledged as a de facto standard in the field~\cite{DBLP:conf/sigsoft/ChenCWYHM23,DBLP:conf/icse/MuhlbauerSKDAS23}. In total, \revision{we run over $17,000$ performance tests, resulting in more than $1,500$ CPU hours.}

\subsection{Extracting Performance-sensitive Dependencies}
\label{sec:find-dep}


Next, we extract the performance-sensitive dependencies, which involve at least one performance-sensitive option. 


\subsubsection{Dependencies by Manual}

To do that by using manual, two authors independently perform the following:

\begin{enumerate}[label=(\alph*)]
     \item The non-performance-sensitive options are filtered out.
    \item Each author analyzed the description of all options remaining. An option is said to have a dependency on the other if its description has mentioned the other option.
    \item In the end, both lead authors combined the list of dependencies identified and discussed among all authors for cross-checking on, e.g., the accuracy, completeness, and relationships between descriptions and related options.
    \item The above process was repeated by two rounds of rigorous review to mitigate bias.
\end{enumerate}

The extracted dependencies represent the information that can be obtained by analyzing the manual. Those dependencies are then further verified by investigating if the two variables can indeed be involved in some logical operation in the relevant code. However, since the dependencies are initially extracted from the manual, they are said to be manual-related. Indeed, as we will show, manual analysis can reveal certain dependencies that are deeply hidden in purely code analysis.

\subsubsection{Dependencies by Code}

To extract dependency information from the code, we again focus on the performance-sensitive options only. We then apply taint analysis using the same aforementioned tools, such that the variable of a performance-sensitive option serves as the source while that of the other option is the sink. Within the results, we check whether there is a call path that involves both variables. If that is the case, we investigate the code segment and examine if both variables are involved in some form of logical operation, such as \texttt{If/Else}, \texttt{Switch}, or \texttt{While} loop. When such a logical relationship can be identified, we say the code indicates that there exists a dependency between the variables/options.

Two authors independently execute the above process, and the final results are discussed and agreed by all authors.

\subsubsection{Why not Testing Systems?}

\revision{For dependencies, in this work, we regard the outcomes from both manual and code analysis above as the ground truth without system testing, since the systems have mostly unclear responses when violating performance-sensitive dependencies. Indeed, Chen et al.~\cite{DBLP:conf/sigsoft/ChenWLLX20} have shown that for up to 74.2\% of the dependencies (we checked that these include many performance-sensitive ones), there are no/incomplete messages in the log from the system under their violations, hence we cannot verify if there is a dependency even when testing on the actual systems.}

\subsubsection{Why not Extract all Dependencies First?}

\revision{We focus explicitly on the performance-sensitive options identified and extract their dependencies, as opposed to extracting all dependencies and then pruning dependencies not containing performance-sensitive options, for two reasons:}

\begin{itemize}
      \item \revision{In our work, extracting the dependencies only on the performance-sensitive options under the scale of our study, which constitutes merely 21\% of the options, has already consumed several months of work per expert (we have two lead authors) merely for one round of analysis (we have two rounds), excluding the discussion time for resolving conflicts with all authors. Therefore, extracting the dependencies for all options and further pruning them down would require significantly increased efforts in the first place, since the number of initial dependencies (pairs of options) to be analyzed can increase considerably along the number of options considered. Further, this would inevitably include many non-performance-sensitive dependencies which we are not interested.}

    \item \revision{Since we are only interested in performance-sensitive dependencies, by which we mean those dependencies that involve at least one performance-sensitive option (which is known), focusing on the performance-sensitive options first and their dependencies would lead to similar results as to extracting all dependencies then prune them down. This is because the known performance-sensitive options are identical between the two processes.}
\end{itemize}




\section{Empirical Study Results}
\label{sec:results}

\subsection{RQ1: Performance-sensitive Options}

\subsubsection{Operationalization}



In comparison to the ground truth obtained by actual system measurement, we report on the true positive rate, false positive rate, and false negative rate of the results obtained by analyzing the manual and the code.

To study the discrepancy of correctly identifying performance-sensitive options by using the manual and the code individually, we use the following patterns observed:

\begin{itemize}
    \item \textbf{Manual Only:} This refers to the true performance-sensitive options (against ground truth) that are found by studying the manual only. \revision{For example, the description form manual for \texttt{max\_execution\_time} option specifies that it determines \textsc{MySQL}'s execution waiting time for SELECT statements. However, it does not involve any control flow related to performance-sensitive operations in the code.}

    \item \textbf{Code Only:} This refers to the true performance-sensitive options (against ground truth) that are identified via analyzing the code but are not noticeable from the manual. For example, the \texttt{server.defer-accept} is merely described in the manual as the listening socket for a TCP protocol \texttt{TCP\_DEFER\_ACCEPT} (``on'' or ``off''). Yet, in the code, it is clear that \texttt{server.defer-accept} directly controls the function \texttt{setsockopt()}, which affects various behaviors of \textsc{Lighttpd}, e.g., network traffic, security, errors, and more.
\end{itemize}

To further understand why analyzing manual or code alone might lead to false positives, we distinguish some patterns:


\begin{itemize}
    \item \textbf{FP}$_1$: The configuration option is completely discarded. However, this might not be noticeable in the manual (since it might not be mentioned) or code (due to complicated call graphs). For example, \texttt{max\_length\_for\_sort\_data} in \textsc{MySQL} is a typical example of \textbf{FP}$_1$. In the manual texts, it clearly indicates that the option is performance-sensitive but it has no impact on the system in the code. \texttt{tls-session-cache-size} from \textsc{Redis} change the default number of TLS sessions cached, which is invoked within the function \texttt{tlsConfigure()} that has a clear impact on the system from the code. However, \texttt{tlsConfigure()} is never triggered.

    \item \textbf{FP}$_2$: The option has limited implication on the performance, i.e., the largest change is less than 1\%. For example, the \texttt{LimitInternalRecursion} in \textsc{Httpd} is identified as performance-sensitive options from both the manual and the code analysis. However, in actual profiling, it caused a maximal performance fluctuation of 0.76\% across the workloads.
    \item \textbf{FP}$_3$: The impact of the configuration option on performance is minimal. That is to say, there is indeed a performance fluctuation after option change, but the maximal magnitude is less than $5\%$. For example, the \texttt{KeepAlive} option in \textsc{Httpd} was deemed to be performance-sensitive from both manual and code, but it merely has a maximum of 2.34\% variation in performance on all workloads.
\end{itemize}

Similarly, for false negative cases, we draw on two patterns:

\begin{itemize}
    \item \textbf{FN}$_1$: There is indeed a significant performance implication. For example. the \texttt{Optimizer\_prune\_level} in \textsc{MySQL} does not exhibit any control relationship with performance-sensitive operations in code. Yet, it significantly impacts the performance by 27\% in the profiling, since its setting is passed to a third-party algorithm.
    \item \textbf{FN}$_2$: The maximal performance implication of the options only marginally beyond $5\%$, e.g., by less than $1\%$, and it is possibly due to randomness in the execution. Yet, for the sake of completeness, we alse classify these as performance-sensitive options. \revision{For example, the \texttt{mapreduce.job.am.node-label-expression} in \textsc{Mapreduce} determines whether range requests are permitted, and it is clearly non-performance-sensitive from the texts in the manual. However, it has non-trivial impacts on performance with a maximum of 5.41\% change across all workloads.}
\end{itemize}


We also discuss the main causes of the above false results.

\begin{table}[t!]
\caption{Ground truth performance-sensitive options identified by measuring the systems.}
\label{tb:rq1.1}
\centering
\adjustbox{max width=\columnwidth}{

\begin{tabular}{llll}
 \toprule

 \textbf{Software} & \textbf{All Options} & \textbf{True Performance-sensitive Options} & \textbf{\%} \\
 
 \midrule

\textsc{Httpd} & 77 & \revision{15} & \revision{19\%} \\
\textsc{Lighttpd} & 62 & \revision{18} & \revision{29\%} \\
\textsc{Nginx} & 79 & \revision{12} & \revision{15\%} \\
\textsc{Redis} & 115 & \revision{36} & \revision{31\%} \\
\textsc{MySQL} & 103 & \revision{26} & \revision{25\%} \\
\textsc{HDFS} & 223 & \revision{58} & \revision{26\%} \\
\textsc{MapReduce} & 218 & \revision{56} & \revision{26\%} \\
\textsc{HBase} & 227 & \revision{35} & \revision{15\%} \\
\textsc{Yarn} & 545 & \revision{86} & \revision{16\%} \\
\textsc{Tomcat} & 45 & \revision{12} & \revision{27\%} \\

 \midrule

\textbf{Total} & 1694 & \revision{354} & \revision{21\%} \\

 \bottomrule
\end{tabular}
}

\end{table}

\begin{table*}[t!]
\centering
\caption{Identified performance-sensitive options (by manual and code) against the ground truth along with the patterns (TP\%, FP\%, and FN\% are the true positive rate, false positive rate, and false negative rate against the ground truth, respectively).}
\label{tb:rq1.2}
\adjustbox{max width=\textwidth}{
\begin{tabular}{llllll|lllll}
 \toprule

 \multirow{2}{*}{\textbf{Software}} & \multicolumn{5}{c|}{\textbf{Manual}} & \multicolumn{5}{c}{\textbf{Code}} \\

 \cmidrule{2-11}

&\textbf{$\#$Options} & \textbf{TP\%} & \textbf{FP\%} & \textbf{FN\%} & \textbf{Manual Only} & \textbf{$\#$Options} & \textbf{TP\%} & \textbf{FP\%} & \textbf{FN\%} & \textbf{Code Only} \\

\midrule

\textsc{Httpd} & 37 & 35\% (13/37) & \revision{39\% (24/62)} & 13\% (2/15) & 20\% (3/15) & 36 & \revision{31\% (11/36)} & \revision{40\% (25/62)} & \revision{27\% (4/15)} & \revision{7\% (1/15)} \\

\textsc{Lighttpd} & 36 & 44\% (16/36) & \revision{45\% (20/44)} & \revision{11\% (2/18)} & \revision{6\% (1/18)} & 38 & 39\% (15/38) & \revision{52\% (23/44)} & \revision{17\% (3/18)} & \revision{0\% (0/18)} \\

\textsc{Nginx} & 52 & \revision{19\% (10/52)} & \revision{63\% (42/67)} & \revision{17\% (2/12)} & \revision{17\% (2/12)} & 47 & \revision{21\% (10/47)} & \revision{55\% (37/67)} & \revision{17\% (2/12)} & \revision{17\% (2/12)} \\

\textsc{Redis} & 81 & \revision{38\% (31/81)} & \revision{63\% (50/79)} & \revision{14\% (5/36)} & \revision{11\% (4/36)} & 75 & \revision{41\% (31/75)} & \revision{56\% (44/79)} & \revision{14\% (5/36)} & \revision{11\% (4/36)} \\

\textsc{MySQL} & 72 & \revision{32\% (23/72)} & \revision{64\% (49/77)} & \revision{12\% (3/26)} & \revision{19\% (5/26)} & 68 & \revision{28\% (19/68)} & \revision{64\% (49/77)} & \revision{27\% (7/26)} & \revision{4\% (1/26)} \\

\textsc{HDFS} & 112 & \revision{44\% (49/112)} & \revision{38\% (63/165)} & \revision{16\% (9/58)} & \revision{7\% (4/58)} & 84 & \revision{57\% (48/84)} & \revision{22\% (36/165)} & \revision{17\% (10/58)} & \revision{5\% (3/58)} \\

\textsc{MapReduce} & 106 & \revision{48\% (51/106)} & \revision{34\% (55/162)} & \revision{9\% (5/56)} & \revision{14\% (8/56)} & 93 & \revision{48\% (45/93)} & \revision{30\% (48/162)} & \revision{20\% (11/56)} & \revision{4\% (2/56)} \\

\textsc{HBase} & 74 & \revision{43\% (32/74)} & \revision{22\% (42/192)} & \revision{9\% (3/35)} & \revision{11\% (4/35)} & 64 & \revision{45\% (29/64)} & \revision{18\% (35/192)} & \revision{17\% (6/35)} & \revision{3\% (1/35)} \\

\textsc{Yarn} & 176 & \revision{45\% (80/176)} & \revision{21\% (96/459)} & \revision{7\% (6/86)} & \revision{10\% (9/86)} & 141 & \revision{51\% (72/141)} & \revision{15\% (69/459)} & \revision{16\% (14/86)} & \revision{1\% (1/86)} \\

\textsc{Tomcat} & 32 & 34\% (11/32) & \revision{64\% (21/33)} & \revision{8\% (1/12)} & \revision{8\% (1/12)} & 26 & \revision{42\% (11/26)} & \revision{45\% (15/33)} & \revision{8\% (1/12)} & \revision{8\% (1/12)} \\

\midrule

\textbf{Total} & 778 & \revision{41\% (316/778)} & \revision{34\% (462/1340)} & \revision{11\% (38/354)} & \revision{12\% (41/354)} & 672 & \revision{43\% (291/672)} & \revision{28\% (381/1340)} & \revision{18\% (63/354)} & \revision{5\% (16/354)} \\

\bottomrule
\end{tabular}
}
\end{table*}

\subsubsection{Findings}

From Table~\ref{tb:rq1.1}, it is clear that the true performance-sensitive options constitute a small proportion of all options---merely \revision{21\%} in general across all the systems. 

\begin{quotebox}
   \noindent
   \textit{\textbf{Finding \thefindingcount:} Only a small set of options, between \revision{$15\%$} and \revision{$31\%$}, can non-trivially impact the performance.}
   \addtocounter{findingcount}{1}
\end{quotebox}

For identifying performance-sensitive options from manual and code, as shown in Table~\ref{tb:rq1.2}, we see that it is easier to identify more options by manual than by code \revision{(778 vs. 672), which makes sense since the naturalness of manual could easily lead to more inclusion.} \revision{However, analyzing the manual does not lead to more true positive samples than using code} (\revision{41\%} vs. \revision{43\%}), meaning that the highly structural nature of code can be more beneficial. Taking the \texttt{tcp\_nopush} in \textsc{Nginx} as an example, this option enables or disables the use of the \texttt{TCP\_NOPUSH} socket option on \texttt{FreeBSDor} of the \texttt{TCP\_CORK} socket option. It is difficult to obtain an accurate answer through manual texts, but in the code, this option is repeatedly invoked in recursive network operations, which strongly implies its performance sensitivity. On the other hand, analyzing the manual can lead to higher false positives while analyzing code is more likely to be false negative prone. However, the benefit of identifying more performance-sensitive options via manual is that it also allows us to find more ones that can only be detected therein, i.e., comparing the \revision{12\%} of \textbf{manual only} to the \revision{5\%} of \textbf{code only}. This suggests the good side of the richer naturalness in the manual: the more comprehensive summary of the interrelations between options can provide more understandable and intuitive information than analyzing code alone. For example, in the source code, there is no performance-sensitive flow for \texttt{optimizer\_prune\_level}, but we can intuitively understand from the manual description that this option controls the third-party heuristic algorithms applied during query optimization, pruning less promising parts of the plan from the optimizer's search space in \textsc{MySQL}.

\begin{table*}[t!]
\centering
\caption{Patterns for false positives/negatives on performance-sensitive options found by analyzing manual and code.}
\label{tb:rq1.3}
\begin{center}
\adjustbox{max width=\textwidth}{
\begin{tabular}{llll|ll||lll|ll}
 \toprule
 \multirow[c]{2}{*}[1pt]{\makecell[c]{\textbf{Software}}} & \multicolumn{5}{c||}{\textbf{Manual}} & \multicolumn{5}{c}{\textbf{Code}} \\

\cmidrule{2-11}
  & \textbf{FP$_{1}$} & \textbf{FP$_{2}$} & \textbf{FP$_{3}$} & \textbf{FN$_{1}$} & \textbf{FN$_{2}$} & \textbf{FP$_{1}$} & \textbf{FP$_{2}$} & \textbf{FP$_{3}$} & \textbf{FN$_{1}$} & \textbf{FN$_{2}$} \\
 \midrule
 \textsc{HTTPD} & 4\% (1/24) & 21\% (5/24) & 75\% (18/24) & \revision{50\% (1/2)} & \revision{50\% (1/2)} & \revision{4\% (1/25)} & \revision{28\% (7/25)} & \revision{68\% (17/25)} & 75\% (3/4) & 25\% (1/4) \\
 \textsc{Lighttpd} & 0\% (0/20) & 5\% (1/20) & 95\% (19/20) & \revision{50\% (1/2)} & \revision{50\% (1/2)} & 0\% (0/23) & \revision{17\% (4/23)} & \revision{83\% (19/23)} & \revision{67\% (2/3)} & \revision{33\% (1/3)} \\
 \textsc{Nginx} & 0\% (0/42) & \revision{55\% (23/42)} & \revision{45\% (19/42)} & \revision{50\% (1/2)} & \revision{50\% (1/2)} & 0\% (0/37) & \revision{59\% (22/37)} & \revision{41\% (15/37)} & \revision{50\% (1/2)} & \revision{50\% (1/2)} \\
 \textsc{Redis} & 4\% (2/50) & 34\% (17/50) & \revision{62\% (31/50)} & \revision{80\% (4/5)} & \revision{20\% (1/5)} & \revision{7\% (3/44)} & \revision{27\% (12/44)} & \revision{66\% (29/44)} & \revision{80\% (4/5)} & \revision{20\% (1/5)} \\
 \textsc{MySQL} & 4\% (2/49) & 29\% (14/49) & 67\% (33/49) & \revision{67\% (2/3)} & \revision{33\% (1/3)} & 4\% (2/49) & 29\% (14/49) & 67\% (33/49) & \revision{57\% (4/7)} & \revision{43\% (3/7)} \\
 \textsc{HDFS} & 0\% (0/63) & 37\% (23/63) & \revision{63\% (40/63)} & \revision{67\% (6/9)} & \revision{33\% (3/9)} & 0\% (0/36) & \revision{42\% (15/36)} & \revision{58\% (21/36)} & \revision{70\% (7/10)} & \revision{30\% (3/10)} \\
 \textsc{MapReduce} & \revision{2\% (1/55)} & \revision{36\% (20/55)} & \revision{62\% (34/55)} & \revision{80\% (4/5)} & \revision{20\% (1/5)} & 2\% (1/48) & \revision{38\% (18/48)} & \revision{60\% (29/48)} & \revision{82\% (9/11)} & \revision{18\% (2/11)} \\
 \textsc{HBase} & 0\% (0/42) & \revision{48\% (20/42)} & \revision{52\% (22/42)} & \revision{67\% (2/3)} & \revision{33\% (1/3)} & 0\% (0/35) & \revision{40\% (14/35)} & \revision{60\% (21/35)} & \revision{67\% (4/6)} & \revision{33\% (2/6)} \\
 \textsc{Yarn} & 0\% (0/96) & \revision{42\% (40/96)} & \revision{58\% (56/96)} & \revision{67\% (4/6)} & \revision{33\% (2/6)} & 0\% (0/69) & \revision{23\% (16/69)} & \revision{77\% (53/69)} & \revision{71\% (10/14)} & \revision{29\% (4/14)} \\
 \textsc{Tomcat} & 0\% (0/21) & \revision{52\% (11/21)} & \revision{48\% (10/21)} & \revision{100\% (1/1)} & \revision{0\% (0/1)} & 0\% (0/15) & \revision{53\% (8/15)} & \revision{47\% (7/15)} & \revision{0\% (0/1)} & \revision{100\% (1/1)} \\
 \midrule
 \textbf{Total} & \revision{1\% (6/462)} & \revision{38\% (174/462)} & \revision{61\% (282/462)} & \revision{68\% (26/38)} & \revision{31\% (12/38)} & \revision{2\% (7/381)} & \revision{34\% (130/381)} & \revision{64\% (244/381)} & \revision{70\% (44/63)} & \revision{30\% (19/63)} \\
 \bottomrule
\end{tabular}
}
\end{center}
\end{table*}

\begin{quotebox}
   \noindent
   \textit{\textbf{Finding \thefindingcount:} Code is more helpful in finding the true performance-sensitive options and with fewer false positives than manual. Yet, the manual can still provide useful information for some that are difficult to identify in code and are less likely to be false negative prone.}
   \addtocounter{findingcount}{1}
\end{quotebox}


As shown in Table~\ref{tb:rq1.3}, clearly, the patterns of false positive for extracting performance-sensitive options from manual and code are of similar characteristics: $\approx \frac{1}{3}$ are due to the fact that the options have limited implications on the performance; $\approx \frac{2}{3}$ are due to noticeable but small implication; a very limited proportion is related to discarded, but have not been clearly marked options. We found that the fundamental cause is also similar: it is mainly due to both the manual and code giving misleading information about options' performance sensitivity. \revision{For example, \texttt{useAsyncIO} in \textsc{Tomcat} is a widely considered performance-sensitive option~\cite{DBLP:conf/nsdi/KarthikeyanNSZB23,DBLP:conf/ic2e/ChibaNHSS19}, since it enables asynchronous IO while both the manual and code indicate such. However, upon the actual profiling, it only has up to 3.62\% performance change over the workloads. }





The patterns of false negatives when analyzing manual and code also differ marginally: ignoring the options that can largely impact performance has only a little higher proportion for code (\revision{70\%}) than manual (\revision{68\%}); while those suspiciously performance-sensitive options are merely a slightly more common pattern for manual (\revision{31\%}) than code (\revision{30\%}). 

Yet, we found that the predominated root causes can differ: 


\begin{itemize}
    \item \textbf{Cause of false negatives in manual analysis:} The texts in the manual have rather vague information related to performance. For instance, from the \textsc{Lighttpd} manual, the description of option \texttt{server.range-requests} is \textit{"defines whether range requests are allowed or not"}, which is highly imprecise. As such, we did not include that as a performance-sensitive option. Yet, in the actual profiling, it can considerably impact the performance.
    

    \item \textbf{Cause of false negatives in code analysis:} The code involves non-standard patterns of executions that affect the performance. For example, in \textsc{MySQL}, \texttt{max\_seeks\_for\_key} is only called in the function \texttt{find\_cost\_for\_ref()}. Within this function, \texttt{max\_seeks\_for\_key} is compared with several similar options and the highest value is returned as the budget, which does not involve our defined performance operations. However, \textsc{MySQL} uses the cost estimate provided by \texttt{find\_cost\_for\_ref()} to choose the most efficient query execution plan.
\end{itemize}

\begin{quotebox}
   \noindent
     \textit{\textbf{Finding \thefindingcount:} The false positives for extracting performance-sensitive options using manual and code are due to similar patterns and causes, but the causes for false negative samples vary even though their patterns are similar.}
   \addtocounter{findingcount}{1}
\end{quotebox}

\subsection{RQ2: Performance-sensitive Dependencies}

\subsubsection{Operationalization}


We found that all the performance-sensitive dependencies identified (via manual or code) are indeed present, from which the unique dependencies for each system are reported, hence there is no false positive. However, there are indeed false negative cases. We observe the following discrepancy patterns when using manual or code:

\begin{itemize}
    \item \textbf{Manual Only:} There exist performance-sensitive dependencies that are only clearly observable in the manual but deeply hidden in the code. For example, in our taint analysis results, there is no code segment where the taint flow between the options \texttt{read\_buffer\_size} and \texttt{select\_into\_buffer\_size} in \textsc{MySQL}. However, the manual provides detailed descriptions of the dependency relationship for them.
    \item \textbf{Code Only:} Similarly, certain performance-sensitive dependencies have not been mentioned in the manual at all but clearly noticeable in the code. For example, in the source code of \textsc{Lighttpd}, if the value of \texttt{server.max-connections} is greater than half of the value for \texttt{server.max-fds}, then the system will log an error. However, such a dependency between the two said options is not mentioned at all in the the manual.
\end{itemize}

For a given dependency, we found two possible patterns depending on the options' type:

\begin{itemize}
    \item \textbf{Mixed Options:} This means one of the options involved is not relevant to performance. For example, in \textsc{MySQL}, the option \textcolor{black}{\texttt{wait\_timeout}} is performance-sensitive while the option \textcolor{black}{\texttt{interactive\_timeout}} is not, but there is a dependency that the former's value should be smaller than that of the latter. 
    

    \item \textbf{Performance Only Options:} This refers to the dependencies wherein both options are performance-sensitive, e.g., the dependency between \texttt{large\_client\_header\_buffers} and \texttt{connection\_pool\_size} in \textsc{Nginx}. 
    
    

\end{itemize}

\revision{In our analysis, a dependency chain involving multiple options is simply a concatenation of different dependencies with two options (providing that they have one option in common),} hence we also analyze the number of options involved in a dependency chain, which is a direct reflection of the dependency of complexity for a system. Again, we showcase how those complex dependency chains, i.e., the number of options involved greater than 3, can be detected via manual and code analysis individually. For example, in \textsc{Redis}, the \texttt{maxmemory} and \texttt{maxmemory-policy} options jointly determine when and which keys need to be deleted to free memory, hence the latter depends on the former. At the same time, option \texttt{lazyfree-lazy-eviction} depends on the policy of \texttt{maxmemory-policy}, and it determines whether the deletion operation will block the processing of commands. 

Since some performance-sensitive dependencies can only be identified in manual not code (and vice versa), we also study why certain dependencies have been missed.



\begin{table}[t!]
\caption{Performance-sensitive dependencies identified. $|\mathcal{C}|$ denotes the number of options involved in a dependency chain.}
\label{tb:rq2.1}
\centering
\adjustbox{max width=\columnwidth}{
\begin{tabular}{lllll}
 \toprule

 \textbf{Software} & \textbf{All Dependencies} & \textbf{Mixed Options} & \textbf{Perf. Only Options} & $|\mathcal{C}|\geq3$ \\
 
 \midrule

\textsc{Httpd} & 4 & \revision{50\% (2/4)} & \revision{50\% (2/4)} & \revision{75\% (3/4)}\\
\textsc{Lighttpd} & 2 & 100\% (2/2) & 0\% (0/2) & 0\% (0/2)\\
\textsc{Nginx} & \revision{6} & \revision{33\% (2/6)} & \revision{66\% (4/6)} & \revision{50\% (3/6)}\\
\textsc{Redis} & 12 & \revision{58\% (7/12)} & \revision{41\% (5/12)} & \revision{42\% (5/12)}\\
\textsc{MySQL} & \revision{8} & \revision{50\% (4/8)} & \revision{50\% (4/8)} & \revision{25\% (2/8)}\\
\textsc{HDFS} & \revision{13} & \revision{76\% (10/13)} & \revision{23\% (3/13)} & \revision{15\% (2/13)}\\
\textsc{MapReduce} & \revision{10} & \revision{70\% (7/10)} & \revision{30\% (3/10)} & \revision{0\% (0/10)}\\
\textsc{HBase} & \revision{7} & \revision{42\% (3/7)} & \revision{57\% (4/7)} & \revision{0\% (0/10)}\\
\textsc{Yarn} & \revision{20} & \revision{60\% (12/20)} & \revision{40\% (8/20)} & \revision{65\% (13/20)}\\
\textsc{Tomcat} & 2 & 0\% (0/2) & 100\% (2/2) & \revision{100\% (2/2)}\\

 \midrule

\textbf{Total} & \revision{84} & \revision{58\% (49/84)} & \revision{41\% (35/84)} & \revision{38\% (32/84)}\\

 \bottomrule
\end{tabular}
}
\end{table}

\begin{table*}[t!]
\centering
\caption{Patterns between performance-sensitive dependencies identified via manual and code. Same formats as Table~\ref{tb:rq2.1}.}
\label{tb:rq2.2}
\adjustbox{max width=\textwidth}{
\begin{tabular}{llllll|lllll}
 \toprule
 \multirow{2}{*}{\textbf{Software}} & \multicolumn{5}{c|}{\textbf{Manual}} & \multicolumn{5}{c}{\textbf{Code}} \\

\cmidrule{2-11}

 & \textbf{$\#$Dependencies} & \textbf{Manual Only} & \textbf{Mixed Options} & \textbf{Perf. Only Options} & $|\mathcal{C}|\geq3$ & \textbf{$\#$Dependencies} & \textbf{Code Only} & \textbf{Mixed Options} & \textbf{Perf. Only Options} & $|\mathcal{C}|\geq3$ \\

\midrule

\textsc{Httpd} & 1 & \revision{0\% (0/4)} & \revision{0\% (0/2)} & \revision{50\% (1/2)} & \revision{33\% (1/3)} & 4 & \revision{75\% (3/4)} & \revision{100\% (2/2)} & 100\% (2/2) & 100\% (3/3) \\
\textsc{Lighttpd} & 0 & 0\% (0/2) & \revision{0\% (0/2)} & 0\% (0/0) & 0\% (0/0) & 2 & \revision{100\% (2/2)} & \revision{100\% (2/2)} & 0\% (0/0) & 0\% (0/0) \\
\textsc{Nginx} & \revision{4} & 33\% (2/6) & 100\% (2/2) & 50\% (2/4) & 100\% (3/3) & 4 & 33\% (2/6) & 50\% (1/2) & 75\% (3/4) & \revision{66\% (2/3)} \\
\textsc{Redis} & \revision{9} & 8\% (1/12) & 57\% (4/7) & 100\% (5/5) & 100\% (5/5) & \revision{11} & 25\% (3/12) & \revision{85\% (6/7)} & 100\% (5/5) & 80\% (4/5) \\
\textsc{MySQL} & \revision{8} & \revision{25\% (2/8)} & 100\% (4/4) & 100\% (4/4) & 100\% (2/2) & 6 & 0\% (0/8) & 75\% (3/4) & 75\% (3/4) & \revision{50\% (1/2)} \\
\textsc{HDFS} & \revision{5} & 7\% (1/13) & \revision{40\% (4/10)} & \revision{33\% (1/3)} & \revision{0\% (0/2)} & 12 & 61\% (8/13) & 90\% (9/10) & 100\% (3/3) & 100\% (2/2) \\
\textsc{MapReduce} & \revision{4} & 10\% (1/10) & \revision{42\% (3/7)} & 33\% (1/3) & 0\% (0/0) & \revision{9} & 60\% (6/10) & 85\% (6/7) & 100\% (3/3) & 0\% (0/0) \\
\textsc{HBase} & 5 & 28\% (2/7) & 66\% (2/3) & 75\% (3/4) & 0\% (0/2) & 5 & 28\% (2/7) & 66\% (2/3) & 75\% (3/4) & \revision{100\% (2/2)} \\
\textsc{Yarn} & \revision{15} & \revision{10\% (2/20)} & 66\% (8/12) & 87\% (7/8) & 76\% (10/13) & \revision{18} & 25\% (5/20) & \revision{100\% (12/12)} & 75\% (6/8) & 92\% (12/13) \\
\textsc{Tomcat} & 2 & 0\% (0/2) & 0\% (0/0) & 100\% (2/2) & 100\% (2/2) & 2 & 0\% (0/2) & 0\% (0/0) & 100\% (2/2) & 100\% (2/2) \\

\midrule
\textbf{Total} & \revision{53} & \revision{20\% (11/53)} & \revision{55\% (27/49)} & \revision{74\% (26/35)} & \revision{71\% (23/32)} & \revision{73} & \revision{42\% (31/73)} & \revision{87\% (43/49)} & \revision{85\% (30/35)} & \revision{87\% (28/32)} \\

 \bottomrule
\end{tabular}
}
\vspace{-0.4cm}
\end{table*}

\subsubsection{Findings}

Table~\ref{tb:rq2.1} shows the general results and patterns for all the systems. We see that the performance-sensitive options can indeed involve dependencies. More than half of those are only be relevant to performance-sensitive options while the remaining half can involve non-performance-sensitive ones. In particular, there is a non-trivial proportion of the chain that covers three or more options (\revision{38\%}), meaning that the complexity of dependency can be considerably high.

\begin{quotebox}
   \noindent
   \textit{\textbf{Finding \thefindingcount:} Among the performance-sensitive dependencies, over half of these involve mixed options while there exist a non-trivial proportion of the chain that involves more than 3 options.}
   \addtocounter{findingcount}{1}
\end{quotebox}

Table~\ref{tb:rq2.2} summarizes the statistics of dependencies that can be identified from manual and code. As can be seen, the code identifies \revision{73} dependencies against the \revision{53} from the manual; since no false positives are involved and there are \revision{84} dependencies in total, this means that the information from the code only misses \revision{11} ones while that of the manual can cause \revision{31} misses. Further, it is much more likely to find dependencies from code that cannot be found in the manual (\revision{42\%}) than the opposed way (\revision{20\%}). The key reason is that, for systems like \textsc{Lighttpd}, there is almost no information about dependencies in the manual. We believe that when the dependencies do not cause serious issues, e.g., crashes, the developers tend to omit their descriptions from the manual.

The same can be reflected in identifying dependencies over the patterns of option types. What we found is that for mixed options, the manual is more likely to miss out on their dependencies. \textcolor{black}{For example, in \textsc{Lighttpd}, \texttt{server.errorlog-use-syslog} is performance-sensitive while \texttt{server.syslog-facility} is not}, and the manual has no information of their dependency. Yet, they clearly have an ``enabling'' dependency in code.


Code also helps to identify most complicated dependency chains (87\%) while manual can only provide clear information on \revision{71\%} of those. For example, in \textsc{Httpd}, the \texttt{KeepAlive} option has a dependency on the \texttt{TimeOut} option, which influences the data flow of the \texttt{KeepAliveTimeOut} option. This dependency chain with three options is not found in the manual but is clearly visible in the code.

\begin{quotebox}
   \noindent
   \textit{\textbf{Finding \thefindingcount:} Code often offers more useful information than manual in identifying performance-sensitive dependencies across the patterns and dependency length.}
   \addtocounter{findingcount}{1}
\end{quotebox}

Since we found no false positive, we focus on the reason that causes false negatives using manual and code individually:

\begin{itemize}
    \item \textbf{Cause of false negatives in manual analysis:} We found that the lack of manual information has led to the \revision{31} false negative cases. Developers tend to omit option dependencies that do not have direct relationships or serious consequences when maintaining the manual, to avoid causing too much trouble for the users.
     \item \textbf{Cause of false negative in code analysis:} For the \revision{11} edge cases where the code analysis has resulted in false negatives, the key reason is due to the depth limitations of taint analysis, which prevent locating all dependencies within the code. When configuration options are called through multiple recursions of function calls, the taint analysis often fails to trace the code, hence leading to most of the false negatives.
\end{itemize}

\begin{quotebox}
   \noindent
   \textit{\textbf{Finding \thefindingcount:} Although code is generally more useful in identifying performance-sensitive dependencies, the manual can still help to resolve some edge cases.}
   \addtocounter{findingcount}{1}
\end{quotebox}

\subsection{RQ3: Usefulness of Current Automated Tools}

\subsubsection{Operationalization}


We now seek to examine the effectiveness of existing automated tools for configuration performance analysis. We choose state-of-the-art tools which relies on the information modality of manual or code independently. Tools for performance-sensitive options inference are:

\begin{itemize}
    \item \textbf{\texttt{SafeTune}~\cite{DBLP:conf/icse/HeJLYZ0WL22}:} As a representative that relies on \textbf{manual}, \texttt{SafeTune} identifies performance-sensitive options by analyzing configuration documentation.
    \item \textbf{\texttt{DiagConfig}~\cite{DBLP:conf/sigsoft/ChenCWYHM23}:} \texttt{DiagConfig} trains a Random Forest to predict whether the given configuration option and its associated code segments are performance-sensitive. As such, it uses \textbf{code} as the main information modality.
\end{itemize}

\revision{Since we did not find readily available tools that can directly extract performance-sensitive dependencies, we use \texttt{GPTuner} and \texttt{CDep}, which are designed to extract any dependencies:}

\begin{itemize}
    \item \textbf{\texttt{GPTuner~\cite{DBLP:journals/pvldb/LaoWLWZCCTW24}}:} \texttt{GPTuner} leverages the understanding capabilities of Large Language Model (LLM) to infer dependencies by reading the texts from the \textbf{manual}. 

    \item \textbf{\texttt{CDep}~\cite{DBLP:conf/sigsoft/ChenWLLX20}:} In \texttt{CDep}, the dependencies are predicted by using rule mining on \textbf{code} based on predefined patterns. 
    
\end{itemize}

\revision{Within the dependencies identified by those tools, we then further extract the performance-sensitive ones according to our ground truth of performance-sensitive options (i.e., at least one option in the pair of options is performance-sensitive). In the evaluation, we only consider those extracted samples against the performance-sensitive dependencies derived from the process in Section~\ref{sec:find-dep}. This mitigates the unfairness caused by the fact that those tools cover any dependency types.}


\revision{Since all tools are cross-project, we use all the manual/code that exists in a target project as the testing data and pre-train them with the original dataset from their authors if possible. Notably, none of our studied systems were studied in \texttt{DiagConfig} hence we used all of their original datasets in training; there are 5 overlapping systems for \texttt{SafeTune} and hence we removed those systems and only used data from the remaining ones (8 systems) in their original datasets to train the model. These prevent data leakage. \texttt{GPTuner} is a special one since it relies on LLM/GPT3.5; as such we have no idea what data it has been using to pre-train the LLM. \texttt{CDep} is a rule-based tool that was designed using general understandings from the systems we studied and hence it is not subject to the data leakage issue in general machine learning-based tools.}

We use recall, precision, and F1 score as the metrics as they are robust to data imbalance~\cite{DBLP:journals/ml/BejDWNW21}. Note that \texttt{DiagConfig} and \texttt{CDep} only work for JAVA-based systems. For all cases, we follow the general ``rule-of-thumb'' that an F1 score of 0.7 or higher means a practically useful configuration predictor~\cite{DBLP:conf/kbse/SharGHSTK23}. We repeat 10 runs of experiments for all stochastic tools.


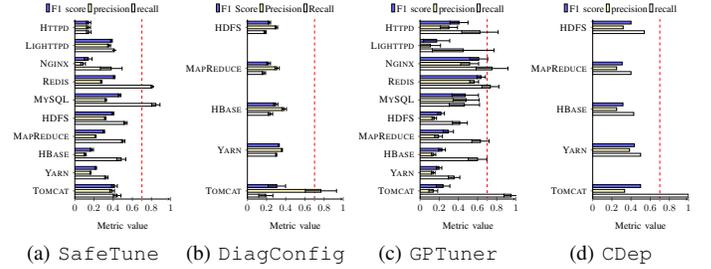
\begin{figure}[t!]
\centering
\subfloat[\texttt{SafeTune}]{
    \begin{tikzpicture}
        \begin{axis}[
            xbar,
            axis x line  = bottom,
            axis y line  = left  ,
             reverse legend,
            bar width=.1cm,
            width=5cm,
            height=8cm,
            enlarge y limits=0.05,
            xlabel={Metric value},
            symbolic y coords={\textsc{Tomcat}, \textsc{Yarn}, \textsc{HBase}, \textsc{MapReduce}, \textsc{HDFS}, \textsc{MySQL}, \textsc{Redis}, \textsc{Nginx}, \textsc{Lighttpd}, \textsc{Httpd}},
            ytick=data,
            xmin=0,
            xmax=1,
             legend style={at={(0.3,1.1)},draw=none,fill=none,anchor=north,legend columns=3},
            legend cell align={left},
            extra x ticks = 0.7,
        extra x tick labels={},
        extra x tick style={grid=major,major grid style={dashed,draw=red}}
        ]

    \addplot+[fill=white, draw=black, error bars/.cd,
              x dir=both,
              x explicit,
              error bar style={black}
    ] plot coordinates {
        (0.146153846,\textsc{Httpd}) +- (0.024325213,0)
        (0.411764706,\textsc{Lighttpd}) +- (0,0)
        (0.38,\textsc{Nginx}) +- (0.113529242,0)
        (0.807692308,\textsc{Redis}) +- (0,0)
        (0.845,\textsc{MySQL}) +- (0.043779752,0)
        (0.528888889,\textsc{HDFS}) +- (0.017529125,0)
        (0.505128205,\textsc{MapReduce}) +- (0.017306373,0)
        (0.484,\textsc{HBase}) +- (0.04788876,0)
        (0.328787879,\textsc{Yarn}) +- (0.017568209,0)
        (0.441666667,\textsc{Tomcat}) +- (0.040253824,0)
    };

    \addplot+[fill=yellow!30, draw=black, error bars/.cd,
              x dir=both,
              x explicit,
              error bar style={black}
    ] plot coordinates {
        (0.142399267,\textsc{Httpd}) +- (0.021459987,0)
        (0.357748538,\textsc{Lighttpd}) +- (0.015662165,0)
        (0.08601901,\textsc{Nginx}) +- (0.024860282,0)
        (0.276490468,\textsc{Redis}) +- (0.00737643,0)
        (0.323063192,\textsc{MySQL}) +- (0.008187825,0)
        (0.317743414,\textsc{HDFS}) +- (0.00800863,0)
        (0.216020266,\textsc{MapReduce}) +- (0.005919184,0)
        (0.108650948,\textsc{HBase}) +- (0.010787279,0)
        (0.163214656,\textsc{Yarn}) +- (0.005719,0)
        (0.389450549,\textsc{Tomcat}) +- (0.024441871,0)
    };

    \addplot+[fill=blue!60, draw=black, error bars/.cd,
              x dir=both,
              x explicit,
              error bar style={black}
    ] plot coordinates {
        (0.144182336,\textsc{Httpd}) +- (0.02272842,0)
        (0.382697961,\textsc{Lighttpd}) +- (0.008891115,0)
        (0.140218152,\textsc{Nginx}) +- (0.040709955,0)
        (0.41190867,\textsc{Redis}) +- (0.00815866,0)
        (0.467292276,\textsc{MySQL}) +- (0.013970988,0)
        (0.396955169,\textsc{HDFS}) +- (0.010652922,0)
        (0.302592447,\textsc{MapReduce}) +- (0.008375981,0)
        (0.177452593,\textsc{HBase}) +- (0.017555192,0)
        (0.218111048,\textsc{Yarn}) +- (0.008741148,0)
        (0.413675214,\textsc{Tomcat}) +- (0.030097123,0)
    };

         \legend{recall, precision, F1 score}
        \end{axis}

    \end{tikzpicture}
}
~\hfill
\subfloat[\texttt{DiagConfig}]{
    \begin{tikzpicture}
        \begin{axis}[
            xbar,
            axis x line  = bottom,
            axis y line  = left  ,
             reverse legend,
             bar width=.1cm,
            width=5cm,
            height=8cm,
            enlarge y limits=0.05,
            xlabel={Metric value},
            symbolic y coords={\textsc{Tomcat}, \textsc{Yarn}, \textsc{HBase}, \textsc{MapReduce}, \textsc{HDFS}},
            ytick=data,
            xmin=0,
            xmax=1,
           legend style={at={(0.3,1.1)},draw=none,fill=none,anchor=north,legend columns=3},
            legend cell align={left},
            extra x ticks = 0.7,
        extra x tick labels={},
        extra x tick style={grid=major,major grid style={dashed,draw=red}}
        ]

    \addplot+[fill=white, draw=black, error bars/.cd,
              x dir=both,
              x explicit,
              error bar style={black}
    ] plot coordinates {
        (0.184482759,\textsc{HDFS}) +- (0.0125,0)
        (0.171428571,\textsc{MapReduce}) +- (0.0181,0)
        (0.24,\textsc{HBase}) +- (0.0232,0)
        (0.3,\textsc{Yarn}) +- (0.0061,0)
        (0.191666667,\textsc{Tomcat}) +- (0.0736,0)
    };

    \addplot+[fill=yellow!30, draw=black, error bars/.cd,
              x dir=both,
              x explicit,
              error bar style={black}
    ] plot coordinates {
        (0.301408451,\textsc{HDFS}) +- (0.0135,0)
        (0.309677419,\textsc{MapReduce}) +- (0.023,0)
        (0.383561644,\textsc{HBase}) +- (0.024,0)
        (0.360335196,\textsc{Yarn}) +- (0.0052,0)
        (0.766666667,\textsc{Tomcat}) +- (0.1626,0)
    };

    \addplot+[fill=blue!60, draw=black, error bars/.cd,
              x dir=both,
              x explicit,
              error bar style={black}
    ] plot coordinates {
        (0.228877005,\textsc{HDFS}) +- (0.013,0)
        (0.220689655,\textsc{MapReduce}) +- (0.0201,0)
        (0.295254833,\textsc{HBase}) +- (0.0221,0)
        (0.327411168,\textsc{Yarn}) +- (0.0056,0)
        (0.306666667,\textsc{Tomcat}) +- (0.0912,0)
    };

    \legend{Recall, Precision, F1 Score}
    \end{axis}
\end{tikzpicture}}
~\hfill
\subfloat[\texttt{GPTuner}]
{
    \begin{tikzpicture}
        \begin{axis}[
            xbar,
            axis x line  = bottom,
            axis y line  = left  ,
             reverse legend,
            bar width=.1cm,
            width=5cm,
            height=8cm,
            enlarge y limits=0.05,
            xlabel={Metric value},
            symbolic y coords={\textsc{Tomcat}, \textsc{Yarn}, \textsc{HBase}, \textsc{MapReduce}, \textsc{HDFS}, \textsc{MySQL}, \textsc{Redis}, \textsc{Nginx}, \textsc{Lighttpd}, \textsc{Httpd}},
            ytick=data,
            xmin=0,
            xmax=1,
            legend style={at={(0.3,1.1)},draw=none,fill=none,anchor=north,legend columns=3},
            legend cell align={left},
              extra x ticks = 0.7,
        extra x tick labels={},
        extra x tick style={grid=major,major grid style={dashed,draw=red}}
        ]

    \addplot+[fill=white, draw=black, error bars/.cd,
              x dir=both,
              x explicit,
              error bar style={black}
    ] plot coordinates {
        (0.625,\textsc{Httpd}) +- (0.1899,0)
        (0.45,\textsc{Lighttpd}) +- (0.3218,0)
        (0.75,\textsc{Nginx}) +- (0.1663,0)
        (0.733333333,\textsc{Redis}) +- (0.0887,0)
        (0.4625,\textsc{MySQL}) +- (0.1577,0)
        (0.415384615,\textsc{HDFS}) +- (0.0776,0)
        (0.63,\textsc{MapReduce}) +- (0.09,0)
        (0.6,\textsc{HBase}) +- (0.0971,0)
        (0.355,\textsc{Yarn}) +- (0.0598,0)
        (0.95,\textsc{Tomcat}) +- (0.075,0)
    };

    \addplot+[fill=yellow!30, draw=black, error bars/.cd,
              x dir=both,
              x explicit,
              error bar style={black}
    ] plot coordinates {
        (0.301204819,\textsc{Httpd}) +- (0.0892,0)
        (0.107142857,\textsc{Lighttpd}) +- (0.1064,0)
        (0.517241379,\textsc{Nginx}) +- (0.0935,0)
        (0.564102564,\textsc{Redis}) +- (0.0464,0)
        (0.480519481,\textsc{MySQL}) +- (0.1354,0)
        (0.148351648,\textsc{HDFS}) +- (0.0236,0)
        (0.19266055,\textsc{MapReduce}) +- (0.0408,0)
        (0.141414141,\textsc{HBase}) +- (0.0225,0)
        (0.137330754,\textsc{Yarn}) +- (0.0191,0)
        (0.137681159,\textsc{Tomcat}) +- (0.0462,0)
    };

    \addplot+[fill=blue!60, draw=black, error bars/.cd,
              x dir=both,
              x explicit,
              error bar style={black}
    ] plot coordinates {
        (0.406504065,\textsc{Httpd}) +- (0.0939,0)
        (0.173076923,\textsc{Lighttpd}) +- (0.1362,0)
        (0.612244898,\textsc{Nginx}) +- (0.0934,0)
        (0.637681159,\textsc{Redis}) +- (0.044,0)
        (0.47133758,\textsc{MySQL}) +- (0.138,0)
        (0.218623482,\textsc{HDFS}) +- (0.0355,0)
        (0.295081967,\textsc{MapReduce}) +- (0.0537,0)
        (0.228882834,\textsc{HBase}) +- (0.0351,0)
        (0.19804742,\textsc{Yarn}) +- (0.0288,0)
        (0.240506329,\textsc{Tomcat}) +- (0.0704,0)
    };

         \legend{recall, precision, F1 score}   
        \end{axis}

    \end{tikzpicture}
}
~\hfill
\subfloat[\texttt{CDep}]{\begin{tikzpicture}
    \begin{axis}[
        xbar,
        axis x line  = bottom,
        axis y line  = left,
        reverse legend,
        bar width=.1cm,
        width=5cm,
        height=8cm,
        enlarge y limits=0.05,
        xlabel={Metric value},
        symbolic y coords={\textsc{Tomcat}, \textsc{Yarn}, \textsc{HBase}, \textsc{MapReduce}, \textsc{HDFS}},
        ytick=data,
        xmin=0,
        xmax=1,
        legend style={at={(0.3,1.1)},draw=none,fill=none,anchor=north,legend columns=3},
        legend cell align={left},
        extra x ticks = 0.7,
        extra x tick labels={},
        extra x tick style={grid=major,major grid style={dashed,draw=red}}
    ]

    \addplot+[fill=white,draw=black
        ] plot coordinates {
        (0.538461538,\textsc{HDFS}) +- (0,0)
        (0.4,\textsc{MapReduce}) +- (0,0)
        (0.428571429,\textsc{HBase}) +- (0,0)
        (0.5,\textsc{Yarn}) +- (0,0)
        (1,\textsc{Tomcat}) +- (0,0) 
    };
    
    \addplot+[fill=yellow!30,draw=black
        ] plot coordinates {
        (0.318181818,\textsc{HDFS}) +- (0,0)
        (0.25,\textsc{MapReduce}) +- (0,0)
        (0.25,\textsc{HBase}) +- (0,0)
        (0.384615385,\textsc{Yarn}) +- (0,0)
        (0.333333333,\textsc{Tomcat}) +- (0,0) 
    };

    \addplot+[fill=blue!60,draw=black
        ] plot coordinates {            
        (0.4,\textsc{HDFS}) +- (0,0) 
        (0.307692308,\textsc{MapReduce}) +- (0,0)
        (0.315789474,\textsc{HBase}) +- (0,0)
        (0.434782609,\textsc{Yarn}) +- (0,0)
        (0.5,\textsc{Tomcat}) +- (0,0) 
    };
    \legend{recall, precision, F1 score}
    \end{axis}

\end{tikzpicture}}
\caption{Effectiveness (mean/deviation) of tools that predict performance-sensitive options (a and b) and configuration dependencies (c and d) over 10 runs. \texttt{SafeTune} and \texttt{GPTuner} are manual-based; \texttt{DiagConfig} and \texttt{CDep} are code-based.}
\label{fig:rq3-1}
\vspace{-0.2cm}
\end{figure}



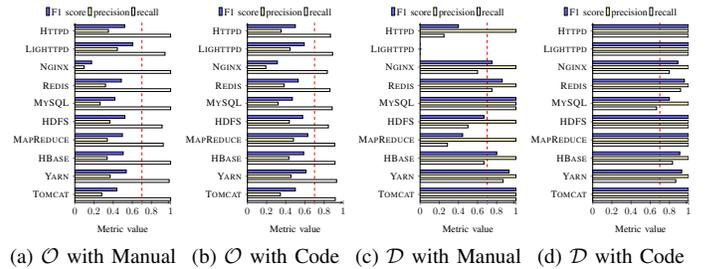
\begin{figure}[t!]
\centering
\subfloat[{$\mathcal{O}$ with Manual}]{\begin{tikzpicture}
    \begin{axis}[
        xbar,
        axis x line  = bottom,
        axis y line  = left,
        reverse legend,
        bar width=.1cm,
        width=5cm,
        height=8cm,
        enlarge y limits=0.05,
        xlabel={Metric value},
        symbolic y coords={\textsc{Tomcat}, \textsc{Yarn}, \textsc{HBase}, \textsc{MapReduce}, \textsc{HDFS}, \textsc{MySQL}, \textsc{Redis}, \textsc{Nginx}, \textsc{Lighttpd}, \textsc{Httpd}},
        ytick=data,
        xmin=0,
        xmax=1,
        legend style={at={(0.3,1.1)},draw=none,fill=none,anchor=north,legend columns=3},
        legend cell align={left},
        extra x ticks = 0.7,
        extra x tick labels={},
        extra x tick style={grid=major,major grid style={dashed,draw=red}}
    ]

    \addplot+[fill=white,draw=black
    ] plot coordinates {
        (1,\textsc{Tomcat}) +- (0,0)
        (0.984848485,\textsc{Yarn}) +- (0,0)
        (1,\textsc{HBase}) +- (0,0)
        (0.923076923,\textsc{MapReduce}) +- (0,0)
        (0.911111111,\textsc{HDFS}) +- (0,0)
        (1,\textsc{MySQL}) +- (0,0)
        (1,\textsc{Redis}) +- (0,0)
        (1,\textsc{Nginx}) +- (0,0)
        (0.941176471,\textsc{Lighttpd}) +- (0,0)
        (1,\textsc{Httpd}) +- (0,0) 
    };
    
    \addplot+[fill=yellow!30,draw=black
    ] plot coordinates {
        (0.28125,\textsc{Tomcat}) +- (0,0)
        (0.369318182,\textsc{Yarn}) +- (0,0)
        (0.337837838,\textsc{HBase}) +- (0,0)
        (0.339622642,\textsc{MapReduce}) +- (0,0)
        (0.366071429,\textsc{HDFS}) +- (0,0)
        (0.263888889,\textsc{MySQL}) +- (0,0)
        (0.320987654,\textsc{Redis}) +- (0,0)
        (0.096153846,\textsc{Nginx}) +- (0,0)
        (0.444444444,\textsc{Lighttpd}) +- (0,0)
        (0.351351351,\textsc{Httpd}) +- (0,0) 
    };

    \addplot+[fill=blue!60,draw=black
    ] plot coordinates {
        (0.43902439,\textsc{Tomcat}) +- (0,0)
        (0.537190083,\textsc{Yarn}) +- (0,0) 
        (0.505050505,\textsc{HBase}) +- (0,0)
        (0.496551724,\textsc{MapReduce}) +- (0,0)
        (0.522292994,\textsc{HDFS}) +- (0,0)
        (0.417582418,\textsc{MySQL}) +- (0,0)
        (0.485981308,\textsc{Redis}) +- (0,0)
        (0.175438596,\textsc{Nginx}) +- (0,0)
        (0.603773585,\textsc{Lighttpd}) +- (0,0)
        (0.52,\textsc{Httpd}) +- (0,0) 
    };
    \legend{recall, precision, F1 score}
    \end{axis}

\end{tikzpicture}}
~\hfill
\subfloat[{$\mathcal{O}$ with Code}]{\begin{tikzpicture}
    \begin{axis}[
        xbar,
        axis x line  = bottom,
        axis y line  = left,
        reverse legend,
        bar width=.1cm,
        width=5cm,
        height=8cm,
        enlarge y limits=0.05,
        xlabel={Metric value},
        symbolic y coords={\textsc{Tomcat}, \textsc{Yarn}, \textsc{HBase}, \textsc{MapReduce}, \textsc{HDFS}, \textsc{MySQL}, \textsc{Redis}, \textsc{Nginx}, \textsc{Lighttpd}, \textsc{Httpd}},
        ytick=data,
        xmin=0,
        xmax=1,
        legend style={at={(0.3,1.1)},draw=none,fill=none,anchor=north,legend columns=3},
        legend cell align={left},
        extra x ticks = 0.7,
        extra x tick labels={},
        extra x tick style={grid=major,major grid style={dashed,draw=red}}
    ]

    \addplot+[fill=white, draw=black
    ] plot coordinates {
        (0.866666667,\textsc{Httpd})
        (0.888888889,\textsc{Lighttpd})
        (0.833333333,\textsc{Nginx})
        (0.861111111,\textsc{Redis})
        (0.884615385,\textsc{MySQL})
        (0.844827586,\textsc{HDFS})
        (0.910714286,\textsc{MapReduce})
        (0.914285714,\textsc{HBase})
        (0.930232558,\textsc{Yarn})
        (0.916666667,\textsc{Tomcat})
    };

    \addplot+[fill=yellow!30, draw=black
    ] plot coordinates {
        (0.351351351,\textsc{Httpd})
        (0.444444444,\textsc{Lighttpd})
        (0.192307692,\textsc{Nginx})
        (0.382716049,\textsc{Redis})
        (0.319444444,\textsc{MySQL})
        (0.4375,\textsc{HDFS})
        (0.481132075,\textsc{MapReduce})
        (0.432432432,\textsc{HBase})
        (0.454545455,\textsc{Yarn})
        (0.34375,\textsc{Tomcat})
    };

    \addplot+[fill=blue!60, draw=black
    ] plot coordinates {
        (0.5,\textsc{Httpd})
        (0.592592593,\textsc{Lighttpd})
        (0.3125,\textsc{Nginx})
        (0.52991453,\textsc{Redis})
        (0.469387755,\textsc{MySQL})
        (0.576470588,\textsc{HDFS})
        (0.62962963,\textsc{MapReduce})
        (0.587155963,\textsc{HBase})
        (0.610687023,\textsc{Yarn})
        (0.5,\textsc{Tomcat})
    };
    \legend{recall, precision, F1 score}
    \end{axis}

\end{tikzpicture}}
~\hfill
\subfloat[{$\mathcal{D}$ with Manual}]{\begin{tikzpicture}
    \begin{axis}[
        xbar,
        axis x line  = bottom,
        axis y line  = left,
        reverse legend,
        bar width=.1cm,
        width=5cm,
        height=8cm,
        enlarge y limits=0.05,
        xlabel={Metric value},
        symbolic y coords={\textsc{Tomcat}, \textsc{Yarn}, \textsc{HBase}, \textsc{MapReduce}, \textsc{HDFS}, \textsc{MySQL}, \textsc{Redis}, \textsc{Nginx}, \textsc{Lighttpd}, \textsc{Httpd}},
        ytick=data,
        xmin=0,
        xmax=1,
        legend style={at={(0.3,1.1)},draw=none,fill=none,anchor=north,legend columns=3},
        legend cell align={left},
        extra x ticks = 0.7,
        extra x tick labels={},
        extra x tick style={grid=major,major grid style={dashed,draw=red}}
    ]

    \addplot+[fill=white, draw=black
    ] plot coordinates {
        (0.25,\textsc{Httpd})
        (0,\textsc{Lighttpd})
        (0.6,\textsc{Nginx})
        (0.75,\textsc{Redis})
        (1,\textsc{MySQL})
        (0.5,\textsc{HDFS})
        (0.285714286,\textsc{MapReduce})
        (0.666666667,\textsc{HBase})
        (0.866666667,\textsc{Yarn})
        (1,\textsc{Tomcat})
    };

    \addplot+[fill=yellow!30, draw=black
    ] plot coordinates {
        (1,\textsc{Httpd})
        (0,\textsc{Lighttpd})
        (1,\textsc{Nginx})
        (1,\textsc{Redis})
        (1,\textsc{MySQL})
        (1,\textsc{HDFS})
        (1,\textsc{MapReduce})
        (1,\textsc{HBase})
        (1,\textsc{Yarn})
        (1,\textsc{Tomcat})
    };

    \addplot+[fill=blue!60, draw=black
    ] plot coordinates {
        (0.4,\textsc{Httpd})
        (0,\textsc{Lighttpd})
        (0.75,\textsc{Nginx})
        (0.857142857,\textsc{Redis})
        (1,\textsc{MySQL})
        (0.666666667,\textsc{HDFS})
        (0.444444444,\textsc{MapReduce})
        (0.8,\textsc{HBase})
        (0.928571429,\textsc{Yarn})
        (1,\textsc{Tomcat})
    };
   \legend{recall, precision, F1 score}
    \end{axis}

\end{tikzpicture}}
~\hfill
\subfloat[{$\mathcal{D}$ with Code}]{\begin{tikzpicture}
    \begin{axis}[
        xbar,
        axis x line  = bottom,
        axis y line  = left,
        reverse legend,
        bar width=.1cm,
        width=5cm,
        height=8cm,
        enlarge y limits=0.05,
        xlabel={Metric value},
        symbolic y coords={\textsc{Tomcat}, \textsc{Yarn}, \textsc{HBase}, \textsc{MapReduce}, \textsc{HDFS}, \textsc{MySQL}, \textsc{Redis}, \textsc{Nginx}, \textsc{Lighttpd}, \textsc{Httpd}},
        ytick=data,
        xmin=0,
        xmax=1,
        legend style={at={(0.3,1.1)},draw=none,fill=none,anchor=north,legend columns=3},
        legend cell align={left},
        extra x ticks = 0.7,
        extra x tick labels={},
        extra x tick style={grid=major,major grid style={dashed,draw=red}}
    ]

    \addplot+[fill=white, draw=black
    ] plot coordinates {
        (1,\textsc{Httpd})
        (1,\textsc{Lighttpd})
        (0.8,\textsc{Nginx})
        (0.916666667,\textsc{Redis})
        (0.666666667,\textsc{MySQL})
        (1,\textsc{HDFS})
        (1,\textsc{MapReduce})
        (0.833333333,\textsc{HBase})
        (0.866666667,\textsc{Yarn})
        (1,\textsc{Tomcat})
    };

    \addplot+[fill=yellow!30, draw=black
    ] plot coordinates {
        (1,\textsc{Httpd})
        (1,\textsc{Lighttpd})
        (1,\textsc{Nginx})
        (1,\textsc{Redis})
        (1,\textsc{MySQL})
        (1,\textsc{HDFS})
        (1,\textsc{MapReduce})
        (1,\textsc{HBase})
        (1,\textsc{Yarn})
        (1,\textsc{Tomcat})
    };

    \addplot+[fill=blue!60, draw=black
    ] plot coordinates {
        (1,\textsc{Httpd})
        (1,\textsc{Lighttpd})
        (0.888888889,\textsc{Nginx})
        (0.956521739,\textsc{Redis})
        (0.8,\textsc{MySQL})
        (1,\textsc{HDFS})
        (1,\textsc{MapReduce})
        (0.909090909,\textsc{HBase})
        (0.928571429,\textsc{Yarn})
        (1,\textsc{Tomcat})
    };
   \legend{recall, precision, F1 score}
    \end{axis}

\end{tikzpicture}}

\caption{Human analysis for performance-sensitive options, i.e., $\mathcal{O}$, (a and b) and their dependencies, i.e., $\mathcal{D}$, (c and d).}
\label{fig:rq3-2}
\vspace{-0.2cm}
\end{figure}



\subsubsection{Findings}

For predicting performance-sensitive options, as from Figures~\ref{fig:rq3-1}a and~\ref{fig:rq3-1}b, we see that manual-based tools like \texttt{SafeTune} would likely have excellent recall but poor precision---this matches with the human analysis of manual in \textbf{RQ1}, where the false positive is more severe than false negatives. As for code-based tools like \texttt{DiagConfig}, the precision is slightly better than the recall. This deviates from the human results form \textbf{RQ1}, because its data-driven nature could parse code information that cannot be captured by human analysts. However, in both cases, we see that the F1 score is still far away from the threshold of 0.7, suggesting that those tools, using either manual or code, are still far from being practically useful in predicting performance-sensitive options.

Figures~\ref{fig:rq3-1}c and~\ref{fig:rq3-1}d illustrate the results for the tools that predict configuration dependencies using different modalities. As can be seen, there is generally poor precision but good recall, regardless of the modality they rely on. This means that false positives are a more serious issue therein, which again deviates from human analysis. As a result, the final F1 score can hardly reach 0.7 on any system. \revision{One possible explanation is that those tools have not been explicitly tailored to cater to the characteristics of performance-sensitive dependencies. For example, \texttt{CDep} is mainly designed based upon the understanding of some manually derived patterns of dependencies, which are relevant to any dependencies type in general. Yet, some of those patterns might not apply to most performance-sensitive dependencies, albeit they can be common in other dependency types. For instance, we found that the \textit{default value pattern}, which represents the dependency wherein the value of an option serves as the default of the other, does not apply to any performance-sensitive dependencies. Often, this is relevant to path-related options, e.g., the value of \texttt{dfs.namenode.name.dir} is the default for \texttt{dfs.namenode.edits.dir} in \textsc{HDFS}. The above can eventually amplify their issue of a high number of false positives for identifying performance-sensitive dependencies when correctly finding the dependencies of other types are no longer of concern, leading to poor precision and negatively impacting their overall performance.}

\begin{quotebox}
   \noindent
   \textit{\textbf{Finding \thefindingcount:} Existing tools for predicting performance-sensitive options and their dependencies tend to be considerably affected by false positives in general, thus can hardly lead to practically useful results.}
   \addtocounter{findingcount}{1}
\end{quotebox}

We also directly compare the results from tools with those of human analysis from \textbf{RQ1} and \textbf{RQ2}. For identifying performance-sensitive options, in Figure~\ref{fig:rq3-1}a and~\ref{fig:rq3-2}a, we see that the F1 score of using tool \texttt{SafeTune} is comparable to that of human analysis when analyzing manual; the latter merely slightly better since understanding natural language also shares similar difficulties as to machine interpretation. However, when comparing the results by analyzing code, we note that human analysis is much superior to the tool, i.e., \texttt{DiagConfig} (Figures~\ref{fig:rq3-1}b and~\ref{fig:rq3-2}b). This is because the code is of complex interrelationships and the learned model in \texttt{DiagConfig} has failed to retain the domain knowledge used in human analysis. Yet, in all cases, neither human analysis nor automated tools have reached the 0.7 F1 score threshold.

For dependencies extraction, clearly the human analysis (Figure~\ref{fig:rq3-2}c and~\ref{fig:rq3-2}d) is significantly better than \texttt{GPTuner} and \texttt{CDep} (Figures~\ref{fig:rq3-1}c and~\ref{fig:rq3-1}d) for the JAVA systems with either modality. Notably, the results of human analysis are mostly beyond the F1 score of 0.7 for all systems.


\begin{quotebox}
   \noindent
   \textit{\textbf{Finding \thefindingcount:} Existing tools are generally far from reaching the level of human analysis, except for identifying performance-sensitive options from manual.}
   \addtocounter{findingcount}{1}
\end{quotebox}



\section{Actionable Insights}
\label{sec:insights}

\textbf{\textit{Finding 1}} clearly suggests that there is only a relatively small proportion of the configuration options that can non-trivially impact the performance. This means that:

\begin{mdframed}
\textbf{Insight 1:} It can be highly beneficial to extract performance-sensitive options within or before tasks such as configuration testing and configuration tuning.
\end{mdframed}

\textbf{\textit{Finding 2}} and 
\textbf{\textit{Finding 3}} imply that, to better identify the performance-sensitive options, it is necessary to jointly investigate the information from both manual and code. Therefore:

\begin{mdframed}
\textbf{Insight 2:} When identifying performance-sensitive options (by hand or with tools), it is necessary to combine the information from both modalities of manual and code.
\end{mdframed}

Yet, the patterns/causes of their false positives are similar but the causes for false negatives can differ:

\begin{mdframed}
\textbf{Insight 3:} When fusing the modality of code and manual for extracting performance-sensitive options, one should not 100\% believe in the information from the manual and code, as it is possibly misleading. In particular, special attention needs to be paid to vague information (for manual) and code that does not follow standard performance-relevant operations.
\end{mdframed}



\revision{The above insight challenges the existing studies (on either manual or code) in which the proposed approach fully trusts the information from the modality, implying the need for some potential confidence-driven approach or some reinforced methods in automated configuration performance learning (even when combining both manual and code).}

\revision{The above also implies that key information might be missing in both modalities, therefore configuration should be better catered to at the software design level. This might include, e.g., standardizing the format between manual/comment and code; clearly annotating the property of configuration options designs, or, at the more fundamental level, ensuring performance regression test is conducted and the results are declared when adding any new configuration options.}

\textbf{\textit{Finding 4}} suggest that when finding performance-sensitive dependencies, \revision{58\%} of the cases involve mixed options. This challenges the belief that dependencies used in, e.g., configuration tuning, are mainly considered for performance-sensitive options only~\cite{DBLP:conf/sigsoft/SiegmundGAK15,DBLP:conf/splc/WeckesserKPMSB18,DBLP:journals/tosem/ChenLBY18}. That said:

\begin{mdframed}
\textbf{Insight 4:} Performance-sensitive dependencies identification should not be specific to performance-sensitive options, as mixed ones are highly possible.
\end{mdframed}

Further, \textbf{\textit{Finding 5}} and 
\textbf{\textit{Finding 6}} suggest that:

\begin{mdframed}
\textbf{Insight 5:} Although it could be acceptable to only rely on the information from the code in identifying performance-sensitive dependencies in general, additionally incorporating manual information can still be useful for finding some edge cases. For manual analysis, the information regarding dependencies might be missing while for the source code, the depth of nested invocations/recursions is critical.
\end{mdframed}

Finally, \textbf{\textit{Finding 7}} and \textbf{\textit{Finding 8}} examine how far are we at automatic performance-sensitive option identification and their dependencies extraction, the results imply that:

\begin{mdframed}
\textbf{Insight 6:} Relying on existing automated tools that leverage either manual or code individually is problematic, and human analysis remains generally far more reliable. Hence it is promising to investigate interactive tools.
\end{mdframed}

\revision{Intuitively, there will be a trade-off between quality (more human analysis) and cost (more tool supports), which is highly case-dependent. Our view is that the inferior accuracy of the tool will have an impact on the downstream tasks benefiting from the configuration performance analysis. For example, if we seek to identify the performance-sensitive options and their dependencies for configuration tuning, then more inaccuracies can lead to unnecessary testing of some options and frequently violated dependencies. Since the configuration tuning is expensive too, in this case, the extra overhead caused by the worse accuracies of a tool might exceed its relative cost-saving.}

\revision{For interactive tools, one possible solution is to incorporate output explainability (e.g., LIME~\cite{DBLP:conf/kdd/Ribeiro0G16} or SHAP~\cite{DBLP:conf/nips/LundbergL17}) into the tool fused with manual and code, based on which humans can then provide their inputs in various forms, e.g., a ``Yes''/``NO'' answer; some categories of confidence, or even weights.}

\section{Threats to Validity}
\label{sec:threats}


The human analysis using manual and code might pose threats to construct validity. To mitigate that, we have followed a systematic protocol involving all authors with multiple rounds of reviews. Yet, \revision{like any empirical study, human errors/oversights are always possible.} \revision{When testing the systems for \textbf{RQ1}, a possible threat can come from the fact that the option is changed one at a time---the standard way that has been followed in many prior works \cite{DBLP:conf/sigsoft/ChenCWYHM23}. Indeed, due to the finite resources and a large number of options/systems to test in our study, fully covering the option interactions is infeasible.} \revision{There are also other forms of configuration beyond this study, e.g., alternative third-party libraries and versions, the understanding of which is promising for future research.}

Threats to internal validity might be related to the parameter setting of the automated tools and the methodology, e.g., the threshold for determining performance sensitivity. In this work, we set the same parameter values and procedures, e.g., the training data for \texttt{DiaConfig}, as used in their prior work. While these serve as a convenient default setting, we cannot guarantee that they are the best status.


To minimize the threats to external validity, we ensure that our study covers a wide range of systems, with diverse manuals and complexity of code. We have also considered tools that leverage different modalities, \revision{as well as using various workloads when testing the systems.} However, studying more systems, workloads, or tools might be more fruitful.




\section{Related Work}
\label{sec:related}

\textit{\textbf{Performance-sensitive Options Identification.}} To identify performance-sensitive options, a common approach has been treating the configurable system as a black box, and hence leveraging data-driven methods in the prediction. For example, some studies have been using empirical observations, e.g,. certain patterns of how the performance-sensitive options influence the performance obtained via qualitative analysis~\cite{DBLP:conf/fast/CaoKZ20,DBLP:conf/hotstorage/KanellisAV20}. Recent work has followed a more white-box approach, where certain artifacts and modalities about configurable systems have been exploited. Among others, He et al.~\cite{DBLP:conf/icse/HeJLYZ0WL22} propose \texttt{SafeTune}, a multi-intent-aware semi-supervised method that analyzes the manuals, aiming to infer the performance-sensitive options.  \texttt{DiagConfig}~\cite{DBLP:conf/sigsoft/ChenCWYHM23} is an approach that also leverages machine learning (random forest). However, it exploits the code invocation chains as part of the features for predicting performance-sensitive options.

\textit{\textbf{Dependency Extraction.}} Existing work on dependency extraction has relied on the systematic analysis of some modalities related to configurable systems.  \texttt{GPTuner}~\cite{DBLP:journals/pvldb/LaoWLWZCCTW24} is an approach that is based on a large language model to parse the manuals for extracting information about dependencies between options. It proposes a prompt integration algorithm to unify the structured view of the refined knowledge. Apart from manual, the other modality that is commonly used is the source code. Zhou et al.~\cite{DBLP:journals/tse/ZhouHXJLWXWLB23} introduce a new tool for multi-configuration error diagnosis by analyzing the dependencies between options in the source code. Similarly, \texttt{Cdep}~\cite{DBLP:conf/sigsoft/ChenWLLX20} detects dependencies between configuration parameters manifested in the code through pattern matching and taint analysis.


\textit{\textbf{Performance Bug Detection.}} There are approaches that rely on either manual or code to detect performance bugs. Among others, \texttt{PracExtractor}~\cite{DBLP:conf/usenix/XiangHY0P20} is a tool that uses natural language processing techniques to automatically extract best practice recommendations from manuals, which serve as an oracle in performance bug detection. \texttt{ECSTATIC}~\cite{DBLP:conf/icse/MordahlZSW23} and \texttt{TAINTMINI}~\cite{DBLP:conf/icse/WangKZYL23} are tools that use taint analysis to parse the flow graph of the code, which detects the potential performance bugs. Yet, these approaches have not covered configuration-related performance.


\textit{\textbf{Related Empirical Studies.}} Empirical studies have been conducted on different aspects of configurable systems~\cite{DBLP:conf/sigsoft/XuJFZPT15,DBLP:conf/icse/ZhangHLL0X21,DBLP:conf/esem/HanY16}. Xu et al.~\cite{DBLP:conf/sigsoft/XuJFZPT15} demonstrate the prevalence of configurability. Zhang et al.~\cite{DBLP:conf/icse/ZhangHLL0X21} investigate how configurations evolve among system versions. However, there has been no empirical study that compares configuration performance analysis using different modalities, i.e., manual or code.

\section{Conclusion}
\label{sec:con}

In this paper, we conduct an extensive empirical study that assesses the usefulness of two modalities, i.e., manual and code, for identifying performance-sensitive options and extracting their dependencies. Through analyzing 10 systems with 1,694 options, 106,798 words in the manual, and 2,859,552 lines-of-code, we reveal several insights that can impact the community of configuration performance analysis. 

Our observations pave the way for vast future research directions, including but not limited to multiple-modal configuration performance analysis and interactive tools thereof.









\section*{Acknowledgement}
This work was supported by a NSFC Grant (62372084) and a UKRI Grant (10054084).

\balance

\bibliographystyle{IEEEtran}
\bibliography{reference}

\vspace{12pt}

\end{document}